\documentclass[12pt]{article}
\usepackage{rayyli} 
\usetikzlibrary{spy,patterns}

\pgfdeclarelayer{fg}    
\pgfsetlayers{main,fg}  

\title{Locality vs Quantum Codes}
\author{Samuel Dai\thanks{Physics Department, Northeastern University. Email: \url{dai.sa@northeastern.edu}.} , Ray Li\thanks{Math \& CS Department, Santa Clara University. Email: \url{rli6@scu.edu}.}}
\date{April 2024}

\begin{document}

\maketitle

\begin{abstract}
    This paper proves optimal tradeoffs between the locality and parameters of quantum error-correcting codes.
    Quantum codes give a promising avenue towards quantum fault tolerance, but the practical constraint of locality limits their quality.
    The seminal Bravyi-Poulin-Terhal (BPT) bound says that a $[[n,k,d]]$ quantum stabilizer code with 2D-locality must satisfy $kd^2\le O(n)$.
    We answer the natural question: for better code parameters, how much ``non-locality" is needed?
    In particular, (i) how long must the long-range interactions be, and (ii) how many long-range interactions must there be?

    We give a complete answer to both questions for all $n,k,d$: above the BPT bound, any 2D-embedding must have at least $\Omega(\#^*)$ interactions of length $\Omega(\ell^*)$, where $\#^*= \max(k,d)$ and $\ell^*=\max\big(\frac{d}{\sqrt{n}}, \big( \frac{kd^2}{n} \big)^{1/4} \big)$.
    Conversely, we exhibit quantum codes that show, in strong ways, that our interaction length $\ell^*$ and interaction count $\#^*$ are asymptotically optimal for all $n,k,d$.
     
    Our results generalize or improve all prior works on this question, including the BPT bound and the results of Baspin and Krishna.
    One takeaway of our work is that, for any desired distance $d$ and dimension $k$, the number of long-range interactions is asymptotically minimized by a good qLDPC code of length $\Theta(\max(k,d))$.
    Following Baspin and Krishna, we also apply our results to the codes implemented in the stacked architecture and obtain better bounds. In particular, we rule out any implementation of hypergraph product codes in the stacked architecture.
\end{abstract}

\newpage
\section{Introduction}

Realizing a quantum computer requires protecting the delicate entanglement of quantum systems that distinguishes quantum computers from classical computers.
Quantum computation needs to be fault-tolerant, regularly correcting the errors due to decoherence and noise.
Quantum error-correcting codes (QECCs) provide a promising avenue towards realizing this fault tolerance, as suggested by the quantum threshold theorem \cite{aharonov1997fault,kitaev1997quantum,knill1998resilient,aliferis2005quantum,shor1996fault}.

There are limitations to implementing quantum codes in practice.
In practical quantum systems, implementing \emph{local} quantum gates --- gates that act on qubits in close spatial proximity --- is easier due to physical constraints on qubit interactions in most hardware architectures.
Consequently, it is desirable to have codes that only use geometrically local gates.
Of particular interest are ``2D-local" codes, where we imagine the code's $n$ qubits are placed in $\mathbb{R}^2$, for example, on a $\sqrt{n}\times \sqrt{n}$ lattice. 
In this setup, for a quantum stabilizer code (the quantum analog of a classical linear code) to use only local quantum gates, the stabilizer group $\mathcal{S}$ must have a set of generators that act on ``close'' qubits, which means that their pairwise distance is a constant $O(1)$.
Some of the most famous classes of quantum codes --- including surface codes \cite{kitaev2003fault,bravyi1998quantum,horsman2012surface} and color codes \cite{bombin2006topological,kubica2015universal} --- enjoy this locality.
However, there are limitations to such ``2D-local" codes.
The seminal Bravyi-Poulin-Terhal (BPT) bound \cite{bravyi2010tradeoffs}, building on the Bravyi-Terhal bound \cite{bravyi2009no}, states that a 2D-local quantum stabilizer code must satisfy $kd^2\le O(n)$, where $n$, $k$, and $d$ denote the length (number of physical qubits), dimension (number of logical qubits), and distance (error tolerance) of the code, respectively.

Although local interactions are easier to implement, many modern quantum architectures offer promising avenues toward supporting non-local interactions, where qubits interact regardless of their physical separation \cite{bergeron2020silicon,bombin2021interleaving,periwal2021programmable,bluvstein2022quantum,monroe2014large,linke2017experimental,murali2020architecting,bravyi2022future,zhong2021deterministic,kurpiers2018deterministic,leung2019deterministic}.
Consequently, it is worthwhile to consider codes without this strict locality constraint.
If we drop the locality constraint, we can construct much better codes.
A recent series of breakthroughs \cite{tillich2014quantum,guth2014quantum,hastings2020fiber,kaufman2021new,evra2022decodable,breuckmann2020balanced,breuckmann2021ldpc,panteleev2020quantum} culminated in \emph{good} codes \cite{panteleev2021asymptotically,leverrier2022quantum} with constant rate $k/n=\Omega(1)$ and constant relative distance $d/n=\Omega(1)$.
These codes are additionally quantum low density parity check (qLDPC), meaning that the stabilizer generators all act on a constant number of qubits.   
These good codes provide no guarantees on the locality.
Inspired by these improvements, much effort has been devoted to implementing these codes and related codes using limited non-locality, to minimize the potential cost of non-local interactions \cite{kovalev2013quantum,hong2023long,eberhardt2024logical, bravyi2024high, berthusen2024partial, berthusen2024toward}. 
This progress suggests the natural question.
\begin{question}
    How much non-locality is needed to implement quantum codes beyond the BPT bound?
    \label{q:main}
\end{question}
As far as we are aware, this question was initiated by \cite{baspin2022quantifying}.
First, we need to quantify what ``non-locality" means. 
The two most natural notions, both of which were considered in \cite{baspin2022quantifying}, were (i) the number of interactions and (ii) length of the interactions.

We give a strong answer to Question~\ref{q:main} that is asymptotically optimal in both measures (i) and (ii) and in \emph{all} parameter regimes.
In the next subsection, we state our main result and compare it to prior work.
In the following subsection, we explain in what ways our result is optimal.
We then discuss related work, and potential future directions.
In the subsequent sections, we prove our main result and its optimality.

\subsection{Main result}

\begin{figure}
\centering
\begin{tikzpicture}[scale=5]
    \draw[->] (0,0) -- (0,1.05);
    \draw[->] (0,0) -- (1.05,0);
    \draw[] (0,0) -- (0,1) -- (1,1) -- (1,0) -- (0,0.5);
    \draw[line width=2pt,purple!10] (1,1) -- (1,0);
    \fill[color=purple!10] (0,1/2)--(0,1) -- (1,1);

    \foreach \i in {4,6,...,10} {
            \tikzmath{\hue = 20+8*\i; \xl = 1.0*\i / 10; \xlprev = 1.0*(\i-2)/10;}
            \draw[thick,blue!\hue] (0,\xl) -- (\xl,\xl) -- (\xl,0);
            \ifthenelse{\i = 10}{
                \node[color=blue!\hue] () at (\xl-0.08,\xl+0.03) {$\#=n^{}$};
            }{
            \node[color=blue!\hue] () at (\xl-0.05,\xl+0.03) {$\#=n^{0.\i}$};
            }
    }
    \node at (1.15,-0.0) () {$\log_n k$};
    \node at (0,1.1) () {$\log_n d$};
    \node at (-.1,0.0) () {$0$};
    \node at (0.0,-.05) () {$0$};
    \node at (-.1,0.5) () {$1/2$};
    \node at (0.5,-.05) () {$1/2$};
    \node at (-.1,1) () {$1$};
    \node at (1,-.05) () {$1$};
    \draw[fill=white] (0,0) -- (0,0.5) -- (1,0) -- (0,0);
    \node[color=blue!20] at (0.3,0.2) {$\#=0$};

    \node (bk22) at (1.1,1.1) {\cite{baspin2022quantifying}*};
    \draw[->] (bk22) to[out=180,in=60] (0.64,0.93);
    \draw[->] (bk22) to[out=300,in=0] (1.01,0.85);
    \node (hmkl23) at (-0.15,0.7) {\cite{hong2023long,fu2024error}};
    \draw[->] (hmkl23) to[out=310,in=120] (-0.01,0.5);
    \node (bpt) at (0.3,0.1) {\cite{bravyi2010tradeoffs}};
    \draw[->] (bpt) to[out=0,in=260] (0.56,0.22);

    \node at (0.5,1.2) {$\#$: Interaction Count};
\end{tikzpicture}
\begin{tikzpicture}[scale=5]
    \draw[->] (0,0) -- (0,1.05);
    \draw[->] (0,0) -- (1.05,0);
    \draw[] (0,0) -- (0,1) -- (1,1) -- (1,0) -- (0,0.5);
      
    \draw[line width=2pt,purple!10] (1,1) -- (1,0);
    \fill[color=purple!10] (0,1/2)--(0,1) -- (1,1);

    \foreach \i in {1,...,5} {
            \tikzmath{\hue = 20+8*\i; \xl = 1.0*\i / 10; \xlprev = 1.0*(\i-1)/10;}
            \draw[thick,blue!\hue] (0,0.5 + \xl) -- (2*\xl,0.5+\xl) -- (1,2*\xl);
            \node[color=blue!\hue] () at (1.15, 2*\xl) {$\ell=n^{0.\i}$};
    }
    \node at (1.15,-0.0) () {$\log_n k$};
    \node at (0,1.1) () {$\log_n d$};
    \node at (-.1,0.0) () {$0$};
    \node at (0.0,-.05) () {$0$};
    \node at (-.1,0.5) () {$1/2$};
    \node at (0.5,-.05) () {$1/2$};
    \node at (-.1,1) () {$1$};
    \node at (1,-.05) () {$1$};
    \node[color=blue!20] at (0.3,0.2) {$\ell=1$};
    \node[rotate=0] at (0.6,0.5) {$\ell=\sqrt[4]{\frac{kd^2}{n}}$};
    \node at (0.2,0.8) {$\ell=\frac{d}{\sqrt{n}}$};

    \node (bk22) at (1.1,1.1) {\cite{baspin2022quantifying}*};
    \draw[->] (bk22) to[out=180,in=60] (0.64,0.93);
    \draw[->] (bk22) to[out=300,in=0] (1.01,0.85);
    \node (hmkl23) at (-0.15,0.7) {\cite{hong2023long,fu2024error}};
    \draw[->] (hmkl23) to[out=310,in=120] (-0.01,0.5);
    \node (bpt) at (0.3,0.1) {\cite{bravyi2010tradeoffs}};
    \draw[->] (bpt) to[out=0,in=260] (0.56,0.22);

    \node at (0.5,1.2) {$\ell$: Interaction Length};
\end{tikzpicture}
\caption{The (asymptotically) optimal interaction count and length for stabilizer codes: A $[[n,k,d]]$ stabilizer code need at least $\Omega(\#)$ interactions of length $\Omega(\ell)$, where $\#$ is plotted on the left and $\ell$ is plotted on the right. 
Above, we plot the counters of $k$ vs. $d$ tradeoffs for various values of the Interaction Count or Interaction Length.
Everywhere, big-$O$ is suppressed for clarity.
The bounds in the purple regions were shown (with lost log-factors) in \cite{baspin2022quantifying}.}
\label{fig:main}
\end{figure}
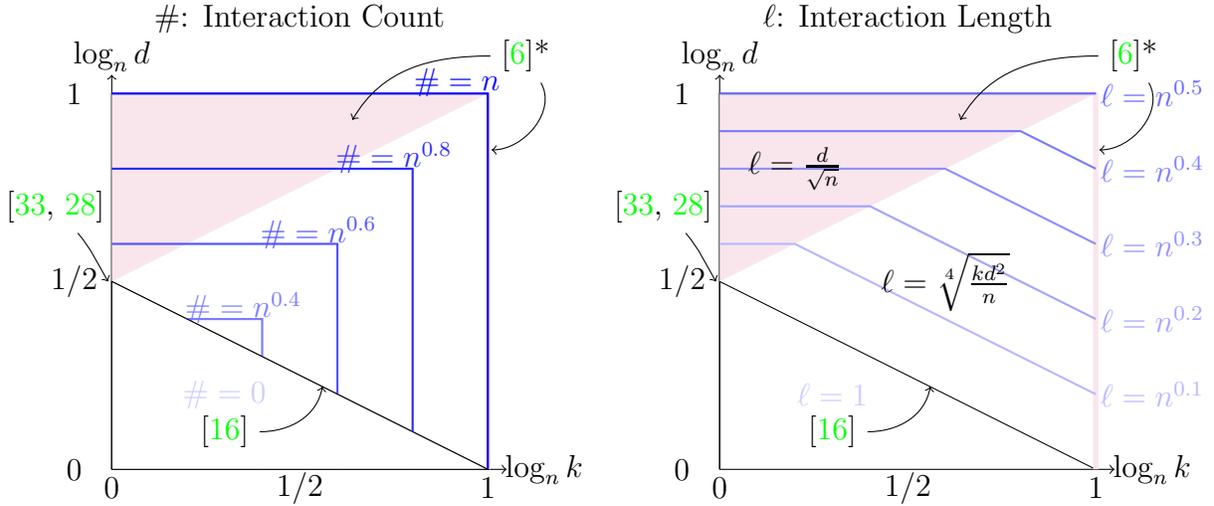

We restrict our attention to stabilizer codes \cite{gottesman1997stabilizer,calderbank1997quantum}, which capture a broad class of quantum codes and are analogous to linear codes in the classical setting.
We expect that our results, like the original BPT bound \cite{bravyi2010tradeoffs}, generalize beyond stabilizer codes.
A \emph{stabilizer code} $\mathcal{Q}$ is specified by a stabilizer group $\mathcal{S}$, an abelian subgroup of the $n$-qubit Pauli group $\mathcal{P}_n$ not containing $-\ssI$. The codewords are the $+1$ eigenstates of the elements in the stabilizer. The stabilizer code can be described by $n-k$ independent generators $\{\ssS_1,...,\ssS_{n-k}\}$, where $k$ is the \emph{dimension} of the code.
The \emph{distance} $d$ is the minimum weight of an error $E$ that maps a codeword in $\mathcal{Q}$ to another codeword.

A \emph{2D-embedding} of a quantum code is a map from its $n$ qubits to $\mathbb{R}^2$ such that any pair of distinct qubits are embedded at least unit distance apart (in Euclidean distance).\footnote{This notion is more general than simply having an embedding on a $\sqrt{n}\times \sqrt{n}$ lattice, and ruling out codes with unusual embeddings is more challenging mathematically.}
For a stabilizer code $\mathcal{Q}$ with a fixed generator set $\{\ssS_1,\dots,\ssS_{n-k}\}$, an \emph{interaction} is a pair of qubits that are both in the support of the same generator $\ssS_i$.
Given a 2D-embedding, the \emph{length} of an interaction is the distance between the embedded qubits. 

Our main result states that, above the BPT bound, any 2D-embedding of a $[[n,k,d]]$ stabilizer code --- a code with length $n$, dimension $k$, and distance $d$ --- must have at least $\Omega(\max(k,d))$ interactions of length $\Omega\big(\max\big(\frac{d}{\sqrt{n}}, \big( \frac{kd^2}{n} \big)^{1/4} \big)\big)$.
For the clarity of exposition, we did not attempt to optimize the implied constants.

\begin{table}
\definecolor{Gray}{gray}{0.85}
\centering
    \begin{tabular}{l|c|c|>{\columncolor{Gray}}c|>{\columncolor{Gray}}c}
        Code name & $k$ & $d$ & \# & $\ell$ \\\hline
        Surface code & 1 & $n^{1/2}$ & --- & --- \\
        2D hyperbolic codes \cite{freedman2002z2,breuckmann2016constructions} & $n$ & $\log(n)$ & $n$ & $\sqrt{\log n}$ \\
        4D hyperbolic codes \cite{guth2014quantum,hastings2013decoding,londe2019golden} & $n$ & $n^\varepsilon$ ($\varepsilon < 0.3$)  & $n$ & $n^{\varepsilon/2}$ \\
        Hypergraph product codes \cite{tillich2014quantum} & $n$ & $n^{1/2}$ & $n$ & $n^{1/4}$\\ 
        Fiber bundle codes \cite{hastings2020fiber} & $n^{3/5}$ & $n^{3/5}$ & $n^{3/5}$ & $n^{1/5}$ \\
        Balanced product codes \cite{breuckmann2020balanced} & $n^{4/5}$ & $n^{3/5}$ & $n^{4/5}$ & $n^{1/4}$ \\
        Codes from HDX \cite{evra2022decodable,kaufman2021new} & $n^{1/2}$
        & $n^{1/2}$ & $n^{1/2}$ & $n^{1/8}$ \\
        Good codes \cite{panteleev2021asymptotically,leverrier2022quantum} & $n$ & $n$ & $n$ & $n^{1/2}$\\
    \end{tabular}
    \caption{Theorem~\ref{thm:main} applied to well-known families of codes. The codes must have $\#$ interactions of length at least $\ell$. Everywhere, $\tilde O$ (constants and logarithmic factors) suppressed for clarity.}
    \label{tab:summary}
\end{table}

\begin{theorem}[Main result]
  There exist absolute constants $c_0,c_1>0$ such that the following holds.
  Any 2D-embedding of a $[[n,k,d]]$ stabilizer code with $kd^2\ge c_1\cdot n$ must\footnote{For silly reasons, the qualifier ``above the BPT bound'' is necessary for our result: the trivial code of dimension $k=n$ with distance 1, has 0 interactions (of any length).} have at least $c_0\cdot \max(k,d)$ interactions of length at least $c_0 \cdot \max\big(\frac{d}{\sqrt{n}}, \big( \frac{kd^2}{n} \big)^{1/4} \big)$.
  \label{thm:main}
\end{theorem}

\paragraph{Comparison to prior work.}
Theorem~\ref{thm:main} generalizes, extends, and/or improves all prior works on this question (see Figure~\ref{fig:main}).
We now discuss the comparison. 

The Bravyi-Terhal bound \cite{bravyi2009no} states that codes with $d\ge \omega(\sqrt{n})$ cannot be 2D-local.
Specifically, it says that a code with $d\ge \omega(\sqrt{n})$ must have at least one interaction of length $\omega(1)$.
The bound on the number of interactions was improved by Baspin and Krishna \cite{baspin2022quantifying} and then (removing log factors) by Hong, Marinelli, Kaufman, and Lucas \cite[Section 2.2]{hong2023long} and Fu and Gottesman \cite[Theorem 72]{fu2024error}. These works showed that a code with $d \ge \omega(\sqrt{n})$ must have at least $\Omega(d)$ interactions of length at least $\omega(1)$.
Up to constant factors, this result matches our result in the special case $d=\omega(\sqrt{n})$ and $k=1$.

The Bravyi-Poulin-Terhal bound \cite{bravyi2010tradeoffs} states that a code with $kd^2\ge \omega(n)$ cannot be 2D-local, i.e., must have at least one interaction of length at least $\omega(1)$.
Theorem~\ref{thm:main} matches this result and additionally obtains a better bound on the number of interactions $\Omega(\max(k,d))$, which, for $kd^2\ge \omega(n)$, is always at least $\omega(n^{1/3})$.

In the parameter regime $d=n^{1/2+\varepsilon}$, Baspin and Krishna \cite{baspin2022quantifying} showed there must be at least $\Omega(d)$ interactions of length $\tilde\Omega(d/\sqrt{n})$, in the specific case that the code is qLDPC.
Theorem~\ref{thm:main} gives the same result when setting $d=n^{1/2+\varepsilon}$, but additionally removes the polylog factors from the bound on the interaction length $\Omega(d/\sqrt{n})$. Moreover, our result holds for general stabilizer codes, not just qLDPC codes. 

In the parameter regime $k=\Theta(n)$, Baspin and Krishna \cite{baspin2022quantifying} showed that, for qLDPC codes, there must be at least $\tilde\Omega(n)$ iterations of length $\tilde\Omega(\sqrt{d})$.
Theorem~\ref{thm:main} gives the same result when setting $k=\Theta(n)$, but additionally remove the polylog factors from both the number of interactions $\Omega(n)$ and interaction length $\Omega(\sqrt{d})$. Again, our result holds for general stabilizer codes. 

For $k$ trading off with $d$, Baspin and Krishna prove that, for qLDPC codes, there are at least $\tilde\Omega(\sqrt{\frac{k}{n}} d)$ interactions of length at least $\tilde\Omega(\sqrt{\frac{kd}{n}})$.
Theorem~\ref{thm:main} improves the interaction count by a factor of $\sqrt{\frac{n}{k}}$ (plus log factors) and the interaction length by a factor of $\sqrt[4]{\frac{n}{k}}$ (plus log factors). Once again, our result holds for general stabilizer codes.

\subsection{Optimality of main result}
\label{ssec:optimal}
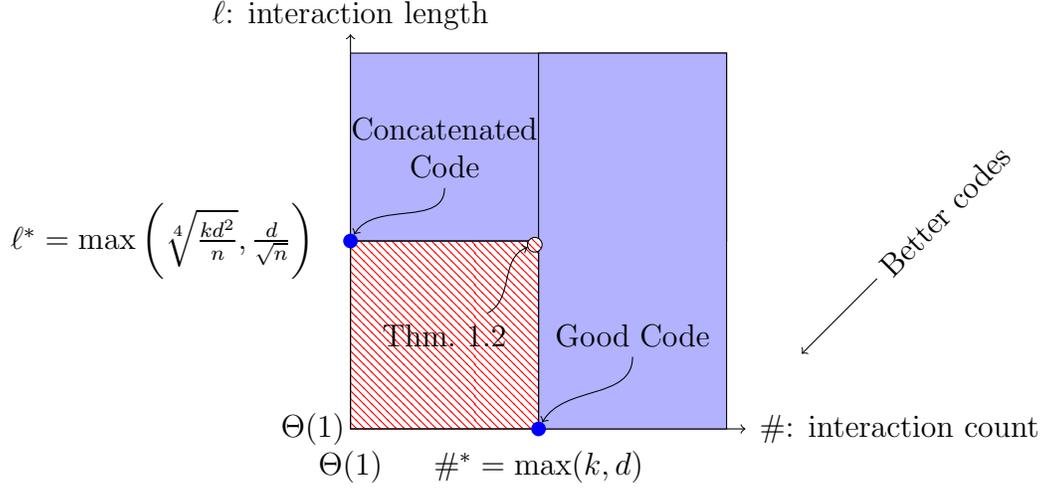
\begin{figure}
\centering
\hypersetup{linkcolor=black}

\begin{tikzpicture}[scale=5]
    \draw[->] (0,0) -- (0,1.05);
    \draw[->] (0,0) -- (1.05,0);
    \draw[fill=blue!30] (0,0.5) rectangle (1,1);
    \draw[fill=blue!30] (0.5,0.00) rectangle (1,1);
    \draw[pattern color=red,pattern=north west lines] (0,0) rectangle (0.5,0.5);
    \node at (1.46,-0.0) () {$\#$: interaction count};
    \node at (0,1.1) () {$\ell$: interaction length};
    \node at (-.1,0.0) () {$\Theta(1)$};
    \node at (0.0,-.1) () {$\Theta(1)$};
    \node at (-0.5,0.5) () {$\ell^*=\max\left( \sqrt[4]{\frac{kd^2}{n}}, \frac{d}{\sqrt{n}} \right)$};
    \node at (0.5,-.1) () {$\#^*=\max(k,d)$};
    \node[preaction={fill,white},pattern color=red,pattern=north west lines,circle,scale=0.5, draw] (bad) at (0.49,0.49) {};
    \node[fill=blue,circle,scale=0.5] (good) at (0.5,0.00) {};
    \node[fill=blue,circle,scale=0.5] (concat) at (0,0.5) {};

    \node[align=center] (concat-label) at (0.25, 0.75) {Concatenated\\Code};
    \node[] (good-label) at (0.75, 0.25) {Good Code};
    \draw[->] (concat-label) to[out=270,in=60] (concat);
    \draw[->] (good-label) to[out=270,in=60] (good);

    \node[] (bad-label) at (0.25, 0.25) {Thm.~\ref{thm:main}};
    \draw[->] (bad-label) to[out=27,in=200] (bad);

    \draw[->] (1.4,0.4) -- (1.2,0.2);
    \node[rotate=45, anchor=west] at (1.4,0.4) {Better codes};

    \node at (0.5,1.3) {\textbf{Interaction Count vs Length}};
\end{tikzpicture}
\caption{Schematic diagram illustrating the optimality of our lower bounds for all $n,k,d$: A point $(\#,\ell)$ represents that there is a code with $O(\#)$ interactions of length $\omega(\ell)$. Blue shaded region is achievable, red lined region is unachievable. Our lower bound shows that $(\#,\ell)$ with $\# \le o(\#^*)$ and $\ell\le o\left(\ell^*\right)$ is impossible, where $\#^*$ and $\ell^*$ are the optimal interaction count and length, respectively, given by Theorem~\ref{thm:main}. There is a construction (good code) with $O(\#^*)$ interactions of any length, and another construction (concatenated construction, Theorem~\ref{thm:construct}) with zero interactions of length $\omega\left(\ell^*\right)$.}
\label{fig:optimal}
\end{figure}

Theorem~\ref{thm:main} is optimal in both the interaction count and interaction length, in the following strong senses (see Figure~\ref{fig:optimal}):

The bound on the number of long interactions is optimal, even if we replace ``long interaction'' with interactions of any length. 
This follows from the existence of good qLDPC codes \cite{panteleev2021asymptotically,leverrier2022quantum}.
For completeness, we work out the connection in Section~\ref{sec:construction}.
\begin{theorem}[Interaction Count is Optimal, \cite{panteleev2021asymptotically,leverrier2022quantum}]
   For all $n,k,d$ (with $k,d\le n$), there are $[[n,\Omega(k),\Omega(d)]]$ qLDPC codes with $O(\max(k,d))$ interactions of any length.
   \label{thm:good}
\end{theorem}
In light of this optimality, one takeaway from our work is that a good qLDPC code \cite{panteleev2021asymptotically,leverrier2022quantum} of length $\Theta(\max(k,d))$ minimizes (up to constant factors) the number of interactions over all codes of dimension at least $k$ and distance at least $d$.

Our interaction length is optimal even if we only guarantee one long interaction.
\begin{theorem}[Interaction Length is Optimal]
   There exists absolute constants $c_0,c_1\ge 1$ such that the following holds.
   For all $n,k,d$ with $k,d\le n$ and $kd^2\ge c_1\cdot n$, there exists a $[[n,\ge k/c_0,\ge d/c_0]]$ quantum stabilizer code with 0 interactions of length at least  $\ell=c_0\cdot\max\big(\frac{d}{\sqrt{n}}, \big( \frac{kd^2}{n} \big)^{1/4} \big)$.
   \label{thm:construct}
\end{theorem}

We establish Theorem~\ref{thm:construct} by concatenating an outer surface code with an inner good code.
A similar concatenation, but in different parameter regimes, and with the outer and inner code swapped, was considered in \cite{pattison2023hierarchical}.
We give the proof in Section~\ref{sec:construction}.

\paragraph{Discussion of Optimality.}
We now elaborate on the strength of our matching constructions, and refer the reader to Figure~\ref{fig:optimal}.
Let $\#^*=\max(k,d)$ and $\ell^* \defeq \max\big(\frac{d}{\sqrt{n}}, \big( \frac{kd^2}{n} \big)^{1/4} \big)$ denote the required interaction count and length, respectively, in Theorem~\ref{thm:main}.

Theorem~\ref{thm:main} implies all possible statements of the following form:
\begin{align}
    \text{Every code has $\Omega(\#)$ interactions of length $\Omega(\ell)$.}\nonumber
\end{align}
Indeed, suppose such a statement were true for some $\#$ and $\ell$.
Theorem~\ref{thm:good} implies that the interaction count satisfies $\# \le O(\#^*)$ (even if we chose a smaller interaction length), and Theorem~\ref{thm:construct} implies that the interaction length satisfies $\ell\le O(\ell^*)$ (even if we chose a smaller interaction count), for all $n,k,d$ above the BPT bound.

We also can see optimality from the construction perspective. First, we remind ourselves of the quantifiers. Theorem~\ref{thm:main} says every code above the BPT bound has ``many'' ``long'' interactions. The negation is that there exist codes with ``few'' ``long'' interactions, so matching constructions should have this form. It is tempting to think that, to show Theorem~\ref{thm:main} is optimal, it suffices to show there are codes with $O(\#^*)$ interactions of length $\Omega(\ell^*)$; this implies optimality in some senses, but our matching constructions are much stronger. Theorem~\ref{thm:good} shows there are codes with $O(\#^*)$ interactions of any length, so it certainly shows there are codes with $O(\#^*)$ interactions of length $\Omega(\ell^*)$.
Theorem~\ref{thm:construct} shows there are codes with 0 interactions of length $\Omega(\ell^*)$, so it certainly shows there are codes with $O(\#^*)$ interactions of length $\Omega(\ell^*)$.

Lastly, to avoid potential confusion, we highlight that there is a more general question for which our main lower bound is not necessarily optimal.
In particular, our result focuses on codes with \emph{one} locality guarantee: ``few'' ``long'' interactions, for one definition of ``few'' and ``long''. 
In general, any 2D-embedded code enjoys a distribution of interaction lengths; as we change the definition of ``long'', the number of ``long'' interactions changes.
Thus, it is natural to wonder whether we could prove stronger lower bounds if we have \emph{multiple} locality guarantees.
One example is codes implemented in the Stacked Architecture, described below in Section~\ref{ssec:stacked}, where we may require, for example \cite{baspin2022quantifying}, $O(n/\ell^2)$ interactions of length $\Omega(\ell)$ for all $\ell$.
Theorem~\ref{thm:main} implies limitations of such codes by focusing on one choice of $\ell$, but perhaps stronger lower bounds are possible if we consider the entire distribution of interaction lengths.
We also suggest another question, inspired by constructions such as bivariate bicycle codes \cite{bravyi2024high} that have many local ($O(1)$-length) interactions and some long interactions.
One could ask, what parameters can be achieved by codes with these two locality guarantees: (i) few interactions of length $\omega(1)$ and (ii) 0 long interactions, for various definitions of ``few'' and ``long''.

\subsection{Application to the Stacked Architecture}
\label{ssec:stacked}
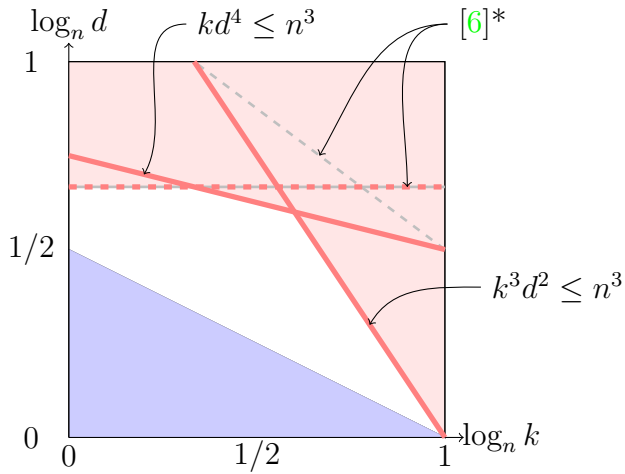
\begin{figure}
    \centering

    \begin{tikzpicture}[scale=5]
    
        \draw[] (0,1/2) -- (1,0);
        \draw[fill, blue!20] (0,1/2) -- (1,0) -- (0,0);
        
        \draw[->] (0,0) -- (0,1.05);
        \draw[->] (0,0) -- (1.05,0);
        \draw[fill=red!10] (0,2/3)--(1/3,2/3)--(3/5,3/5)--(1,0)--(1,1)--(0,1);
        \draw[] (0,0) -- (0,1) -- (1,1) -- (1,0);
        \draw[line width=1pt,dashed,gray!50] (1,1/2) -- (1/3,1);
        \draw[line width=1pt,gray!50] (0,2/3)--(1,2/3);
        \draw[line width=2pt,red!50] (1,0)--(1/3,1);
        \draw[line width=2pt,red!50] (0,3/4)--(1,1/2);
        \draw[line width=2pt,dashed,red!50] (0,2/3)--(1,2/3);
    
        \node at (1.15,-0.0) () {$\log_n k$};
        \node at (0,1.1) () {$\log_n d$};
        \node at (-.1,0.0) () {$0$};
        \node at (0.0,-.05) () {$0$};
        \node at (-.1,0.5) () {$1/2$};
        \node at (0.5,-.05) () {$1/2$};
        \node at (-.1,1) () {$1$};
        \node at (1,-.05) () {$1$};

        \node (bk22) at (1.1,1.1) {\cite{baspin2022quantifying}*};
        \draw[->] (bk22) to[out=180,in=60] (2/3,3/4);
        \draw[->] (bk22) to[out=180,in=90] (0.9,2/3);
                
        \node (kd4) at (0.5,1.1) {$kd^4\le n^3$};
        \draw[->] (kd4) to[out=180,in=60] (0.2,0.7);

        \node (k3d2) at (1.3,0.4) {$k^3d^2\le n^3$};
        \draw[->] (k3d2) to[out=180,in=60] (0.8,0.3);
        
        \node at (0.5,1.4) {\textbf{Achievable qLDPC parameters for Stacked Architecture}};
    \end{tikzpicture}
    \caption{Achievable qLDPC parameters for the stacked architecture as defined in \cite{baspin2022quantifying} Big-$O$ is suppressed for clarity. Red regions are not achievable. Blue region (below BPT bound) is achievable by surface code, and we are not aware of any better constructions. Our bounds imply $d\le O(n^{2/3})$, $kd^4\le O(n^3)$, and $k^3d^2\le O(n^3)$. Baspin and Krishna \cite{baspin2022quantifying} showed $d\le O(n^{2/3}\log^{2/3} n)$ and $k^3d^4\le O(n^5\log^4n)$.}
    \label{fig:stacked}
\end{figure}

We apply our main result to the \emph{stacked architecture}, a model for implementing quantum codes, particularly qLDPC codes, so that most of the interactions are local.
To illustrate our results, we focus on the description of the stacked architecture given in \cite{baspin2022quantifying}, and then discuss implications for a more general definition considered in later works \cite{berthusen2024partial, berthusen2024toward}.
For a qLDPC code implemented in the stacked architecture, the stabilizer generators are partitioned into layers, where layer $i$ consists of generators acting on a constant number of qubits in a ball of radius $O(2^i)$.
In the definition in \cite{baspin2022quantifying}, these balls mostly partition the plane.
Thus, for qLDPC codes implemented in the stacked architecture, layer $i$ consists of $O(n/2^{2i})$ interactions of length $O(2^i)$.
Summing a geometric series, it follows that, for any $\ell$, there are at most $O(n/\ell^2)$ interactions of length $\ge \ell$.
Combining this property --- $O(n/\ell^2)$ interactions of length $\ge\ell$ --- directly with Theorem~\ref{thm:main} produces the following result.

\begin{corollary}
    An $[[n,k,d]]$ qLDPC code implemented in the stacked architecture as defined in \cite{baspin2022quantifying} satisfies $d\le O(n^{2/3})$ and $kd^2\cdot \max(k,d)^2 \le O(n^3)$.
    \label{cor:stacked}
\end{corollary}
\begin{proof}
    To see the first bound $d\le O(n^{2/3})$, pick $\ell = c_0\cdot \frac{d}{\sqrt{n}}$ where $c_0$ is given by Theorem~\ref{thm:main}.
By Theorem~\ref{thm:main}, we must have $O(n/\ell^2) \ge \Omega(\max(k,d)) = \Omega(d)$. Plugging in $\ell$ and rearranging gives the desired bound.
The second bound $kd^2\cdot \max(k,d)^2 \le O(n^3)$ follows similarly. Choose $\ell=c_0\cdot \sqrt[4]{\frac{kd^2}{n}}$.
Again, by Theorem~\ref{thm:main}, we have $O(n/\ell^2)\ge \Omega(\max(k,d))$. Plugging in $\ell$ and rearranging gives the desired bound.
\end{proof}
Figure~\ref{fig:stacked} compares our bounds with those of Baspin and Krishna, who showed $d\le O(n^{2/3}\log n)$ and $k^3d^4\le O(n^5\log^4 n)$.
Our bound rules out more classes of qLDPC codes that can be implemented in the stacked architecture. In particular,  after \cite{baspin2022quantifying}, it remained possible that hypergraph product codes $(k=\Theta(n), d= \Theta(n^{1/2}))$ \cite{tillich2014quantum} and balanced product codes $k=\tilde\Theta(n^{4/5})$ and $d=\tilde\Theta(n^{3/5})$) \cite{breuckmann2020balanced} could be implemented in the stacked architecture.
Our bound rules out these possibilities. 

In fact, we can make a stronger statement on the distribution $f(\ell)$ of the stabilizer generator interaction lengths $\ell$. Theorem~\ref{thm:main} implies that any embedding of a hypergraph product code must have $\Omega(n)$ interactions of length $\Omega(n^{1/4})$, which actually rules out implementing hypergraph product codes in the stacked architecture for \emph{any} locality distribution $f(\ell)$ that decreases with $\ell$ (not just $f(\ell)=O(n/\ell^2)$, see \cite{berthusen2024partial,berthusen2024toward} for this more general notion). This corroborates the observation made by Berthusen and Gottesman \cite{berthusen2024partial}, where they ``[made] some effort...to find specific embeddings [of hypergraph product codes in the stacked architecture]...that yielded good generator size distributions; however, the resulting distributions instead often favored mid-sized generators". 

On the other hand, it still remains possible that the following codes can be implemented in the stacked architecture: fiber bundle codes ($k,d=\tilde\Theta(n^{3/5})$) \cite{hastings2020fiber}, codes based on high-dimensional expanders (with $k,d=\tilde\Theta(\sqrt{n})$) \cite{evra2022decodable,kaufman2021new}, 2D hyperbolic codes (with $k = \Theta(n)$ and $d = O(\log(n))$) \cite{freedman2002z2,breuckmann2016constructions}, and 4D hyperbolic codes (with $k = \Theta(n)$ and $d = O(n^\varepsilon)$ for $\varepsilon < 0.3$) \cite{guth2014quantum,hastings2013decoding,londe2019golden}. As described in \cite{berthusen2024partial}, a fault-tolerant scheme where certain layers of generators are measured more frequently than others could potentially be applicable to these codes. 

\subsection{Related Work}

Several generalizations and extensions of the Bravyi-Terhal bound and Bravyi-Poulin-Terhal bound are known. The same bounds state that an $[[n,k,d]]$ quantum code that is local in $D$ dimensions must satisfy $d\le O(n^{(D-1)/D})$ \cite{bravyi2009no} and $kd^{2/(D-1)} \le O(n)$ \cite{bravyi2010tradeoffs}.
Several recent works by \cite{portnoy2023local,williamson2023layer,lin2023geometrically,li2024transform} constructed codes attaining these higher-dimensional BT and BPT bounds. 
Delfosse showed \cite{delfosse2013tradeoffs} that 2D hyperbolic surface codes satisfy $kd^2\le O(n\log^2 k)$. 
Baspin, Guruswami, Krishna, and Li \cite{baspin2022connectivity,baspin2023improved} have also generalized these bounds to broader classes of codes. These classes are defined by the codes' \emph{connectivity graphs}, the graph whose vertices are the code's qubits and whose edges are the interactions.
Flammia, Haah, Kastoryano, and Kim \cite{flammia2017limits} generalized the BPT bound to a notion of approximate quantum error-correcting codes.
Related to these works, Kalachev and Sadov \cite{kalachev2022linear} proved a generalization of the Cleaning Lemma \cite{bravyi2009no}, the central tool underlying all of these results (including this work\footnote{We do not explicitly state the Cleaning Lemma, but we use it implicitly when we use the Union Lemma and Expansion Lemma (Lemma~\ref{lem:correctable}).}).

Beyond the code dimension and distance, several works have studied how locality limits the circuits that correct the errors in quantum codes. Delfosse, Beverland, and Tremblay \cite{delfosse2021bounds} established limitations of syndrome extraction circuits obtained from 2D local gates. Baspin, Fawzi, and Shayeghi \cite{baspin2023lower} proved tradeoffs between the logical error rate and parameters of the syndrome extraction circuit, for (not necessarily local) codes on architectures with local operations. 

Our construction showing the optimality of the interaction length is obtained by concatenating an outer surface code \cite{kitaev1997quantum} with an inner good qLDPC code \cite{panteleev2021asymptotically,leverrier2022quantum}.
A similar construction was considered  by Pattison, Krishna, and Preskill \cite{pattison2023hierarchical}.
Their work concatenated the codes in the other direction, using an outer qLDPC code with an inner surface code, which showed that the resulting code admits local syndrome extraction circuits with good depth and width.

\subsection{Discussion}

We summarize our contributions and takeaways from our work, and also highlight some directions for future work.
\begin{enumerate}
    \item We show lower bounds on the interaction count and interaction length for long-range interactions in quantum codes beyond the BPT bound, and we show these lower bounds are optimal in strong senses. 
    This completely answers the posed question about the tradeoff between locality and the parameters of quantum stabilizer codes.

    \item Our codes imply interesting nonlocality bounds for many natural families of codes, as illustrated in Table~\ref{tab:summary}.
    
    \item For any desired code dimension and distance, the number of interactions is minimized by a good qLDPC code \cite{panteleev2021asymptotically,leverrier2022quantum}  minimizes (up to constant factors).
Much effort has been devoted to implementing codes using limited non-locality to minimize the potential cost of non-local interactions, possibly by sacrificing code parameters \cite{kovalev2013quantum,hong2023long,eberhardt2024logical, bravyi2024high, berthusen2024partial, berthusen2024toward}.
Our results imply that it is not possible to asymptotically reduce the number of interactions by sacrificing code parameters.

\item Our main result, Theorem~\ref{thm:main}, immediately implies stronger limitations on codes implemented in the stacked architecture. We show that hypergraph product codes cannot be implemented in the stacked architecture, even for general locality distributions, and also that balanced product codes cannot be implemented in the stacked architecture according to the model in \cite{baspin2022quantifying}.

\item Our main result, Theorem~\ref{thm:main} holds in general for stabilizer codes. Can we prove a stronger bound for qLDPC codes? We note that the optimality of the Interaction Count is achieved with a qLDPC code, but Theorem~\ref{thm:construct}, which gives the optimality of the Interaction Length, does not construct a qLDPC code. Thus, it is possible that, for qLDPC codes, we could improve Theorem~\ref{thm:main} to say that a qLDPC code needs at least $\Omega(\max(k,d))$ interactions of length $\Omega(\ell)$, for some larger $\ell > \max\big(\sqrt[4]{\frac{kd^2}{n}}, \frac{d}{\sqrt{n}}\big)$. We leave this for future work. 

\item As discussed in Section~\ref{ssec:optimal}, our lower bound is not necessarily optimal when we consider multiple locality guarantees. For example, can we prove stronger limitations of locality for codes in the stacked architecture? Can we find matching constructions? 

\item For clarity and brevity, we focused on stabilizer codes 2-dimensional embeddings, but we expect our bounds can be generalized beyond these settings. Just as the BPT bound \cite{bravyi2010tradeoffs} holds for non-stabilizer codes, we surmise that our techniques can be generalized to non-stabilizer codes as well. While we focus on the most practically relevant setting of 2-dimensional embeddings, we expect that our techniques can be extended to $D$-dimensional codes., just as the BPT bound holds for higher dimensional codes 

\end{enumerate}

\subsection{Organization of the Paper}
In Section~\ref{sec:overview} we provide an overview of the proof of our main result, Theorem~\ref{thm:main}.
In Section~\ref{sec:prelims}, we establish some preliminaries.
In Section~\ref{sec:proof}, we prove Theorem~\ref{thm:main}.
In Section~\ref{sec:construction}, we prove Theorem~\ref{thm:good} and Theorem~\ref{thm:construct}, which establish optimality of tHeorem~\ref{thm:main}.

In Section~\ref{sec:proof}, we divide the main result into two parts. 
\begin{theorem}[Main result, part 1]
  There exists absolute constants $c_0,c_1>0$ such that the following holds.
  Any 2D-embedding of a $[[n,k,d]]$ stabilizer code with $kd^2\ge c_1\cdot n$ must have at least $c_0\cdot d$ interactions of length at least $c_0\cdot \frac{d}{\sqrt{n}}$.
  \label{thm:main-2}
\end{theorem}
\begin{theorem}[Main result, part 2]
  There exists absolute constants $c_0,c_1>0$ such that the following holds.
  Any 2D-embedding of a $[[n,k,d]]$ stabilizer code with $kd^2\ge c_1\cdot n$  must have at least $c_0\cdot\max(k,d)$ interactions of length $c_0\cdot \big(\frac{kd^2}{n} \big)^{1/4}$.
  \label{thm:main-3}
\end{theorem}
It is easy to see that these two results imply the main result: Take $c_0^{\text{Thm~\ref{thm:main}}}=\min(c_{0}^{\text{Thm~\ref{thm:main-2}}}, c_{0}^{\text{Thm~\ref{thm:main-3}}})$ and $c_1^{\text{Thm~\ref{thm:main}}}=\max(c_1^\text{Thm~\ref{thm:main-2}}, c_1^\text{Thm~\ref{thm:main-3}})$. Theorem~\ref{thm:main-2} gives Theorem~\ref{thm:main} when $d\ge \sqrt{kn}$, and Theorem~\ref{thm:main-3} establishes Theorem~\ref{thm:main} when $d\le \sqrt{kn}$.

\section{Proof Overview}
\label{sec:overview}

\subsection{Technical comparison with \cite{baspin2022quantifying}.}
To start our overview, we first highlight, at the highest level, the technical difference between our approach and the approach in \cite{baspin2022quantifying}, and how our approach obtains better bounds.
\cite{baspin2022quantifying} explicitly leverages the \emph{connectivity graph} of a stabilizer code (see Remark~\ref{rem:graph} for definition).
They transfer the 2D-embedding of the code into graph theoretic properties of the \emph{connectivity graph}, apply graph theoretic tools to the connectivity graph by analyzing the \emph{separator profile} of the graph, and finally convert these graph theoretic results back to the 2D-geometry, giving bounds on the interaction length and count for such codes.
By contrast, our work sticks to the 2D-geometry without passing to the graph interpretation.
Intuitively, it seems difficult to analyze the graph problem as tightly as the corresponding geometric problem (compare, for example, \cite{baspin2022connectivity,baspin2023improved} with \cite{bravyi2010tradeoffs}), and we indeed get a tighter and more general analysis by working directly with the geometry of the problem.

\subsection{Toy version: grid, one long interaction}

We now give some intuition for the strategies used in the formal proof.
We start by proving a toy version of Theorem~\ref{thm:main}, and progressively build up the ideas in the full proof. For the toy version, we make two simplifications from Theorem~\ref{thm:main}: (i) we only prove the existence of \emph{one} long interaction, and (ii) we assume the embedding is a $\sqrt{n}\times \sqrt{n}$ lattice. For simplicity, we also focus on Theorem~\ref{thm:main-3} (the case $d\le \sqrt{kn}$), as this case contains all the new ideas in the proof.
\begin{proposition}
  Any embedding of a $[[n,k,d]]$ qLDPC code on the $\sqrt{n}\times \sqrt{n}$ lattice must have at least one interaction of length $\Omega\big(\big(\frac{kd^2}{n} \big)^{1/4}\big)$.
  \label{pr:main}
\end{proposition}

Suppose our quantum code has a 2D-embedding on a $\sqrt{n}\times \sqrt{n}$ lattice with \emph{no} interactions of length at least $\ell$, for some $\ell$.
We show $\ell\ge \Omega\big(\big(\frac{kd^2}{n} \big)^{1/4}\big)$, implying Proposition~\ref{pr:main}.
Our code is essentially 2D-local, except all interactions are now contained inside balls of radius $\ell$. We aim to partition the qubits into three sets $A,B,C$, where $A,B$ are \emph{correctable}, and $C$ is small.
Intuitively, a set of qubits is correctable if the quantum code can correct the erasure of those qubits.
We construct $A$ and $B$ using four properties of correctable sets given in \cite{bravyi2009no,bravyi2010tradeoffs} (Lemma~\ref{lem:correctable}): Subset Closure (subsets of correctable sets are correctable), Distance Property (sets of less than $d$ qubits are correctable), Union Lemma (the union of disjoint, non-interacting correctable sets is correctable), and the Expansion Lemma (correctable sets are closed under adding a correctable boundary).
We show that we can find large enough $A$ and $B$ so that $|C|\le O(\frac{n\ell^4}{d^2})$. By a standard lemma on correctable sets by Bravyi, Poulin, and Terhal \cite{bravyi2010tradeoffs} (Lemma~\ref{lem:bptabc}), this partition implies $k\le|C|\le O(\frac{n\ell^4}{d^2})$, and rearranging gives our bound on $\ell$.

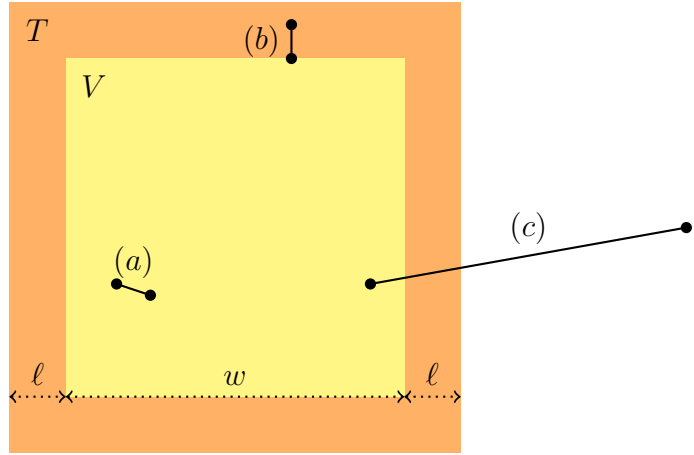
\begin{figure}
    \centering
    \begin{tikzpicture}[scale=1.5]
        \draw[draw=none,fill=orange!60] (-2.,-2.) rectangle (2,2);
        \draw[draw=none,fill=yellow!60] (-1.5, -1.5) rectangle (1.5, 1.5);
        \node at (-1.25,1.25) {$V$};
        \node at (-1.75, 1.75) {$T$};

        \draw[<->, dotted, thick] (-1.5, -1.5) -- (1.5, -1.5) node [midway, above] {$w$};
        \draw[<->, dotted, thick] (-2., -1.5) -- (-1.5, -1.5) node [midway, above] {$\ell$};
        \draw[<->, dotted, thick] (1.5, -1.5) -- (2., -1.5) node [midway, above] {$\ell$};

        \node at (-1.05,-.5) [circle,fill,inner sep=1.5pt]{};
        \draw[thick] (-1.05, -.5) -- (-.75,-.6) node [midway, above] {$(a)$};
        \node at (-.75,-.6) [circle,fill,inner sep=1.5pt]{};

        \node at (.5, 1.5) [circle, fill, inner sep=1.5pt]{};
        \draw[thick] (.5, 1.5) -- (.5, 1.8) node [midway, left]{$(b)$};
        \node at (.5, 1.8) [circle, fill, inner sep=1.5pt]{};

        \node at (1.2, -.5) [circle, fill, inner sep=1.5pt]{};
        \draw[thick] (1.2, -.5) -- (4, 0) node [midway, above]{$(c)$};
        \node at (4, 0) [circle, fill, inner sep=1.5pt]{};
    \end{tikzpicture}
    \caption{Illustration of the growing-the-square process. The yellow internal square $V$ is the small correctable set with side length $w$, the orange external square $T$ is the larger set grown from it with side length $w + 2\ell$. The labeled edges are (a) an interaction between two qubits in $V$, (b) a short interaction between a qubit in $V$ and a qubit in $T$, and (c) a long interaction between a qubit in $V$ and a qubit outside of $T$. }
    \label{fig:growing-the-square}
\end{figure}

To construct large correctable sets, we use the Union and Expansion Lemmas (Lemma~\ref{lem:correctable}) to increase the size of a small correctable square. First, we note that a square with dimensions $\sqrt{d}\times\sqrt{d}$ is correctable by the distance property of Lemma $\ref{lem:correctable}$. Now, since any qubit interacting with a qubit in the square must lie within a ball of radius $\ell$ of the square, we note that a larger square of dimensions $\sqrt{d}+2\ell$ sharing the same center as the original square must contain all such interacting qubits. As long as the boundary of this region contains less than $d$ qubits, it is correctable. Then the Expansion Lemma says that the larger square is correctable as well. We can repeat this process, each time starting with a correctable square of dimensions $(w-2\ell)\times (w-2\ell)$, and then using the Expansion Lemma to find a correctable square of dimensions $w \times w$. This works as long as the boundary of the $w \times w$ square contains less than $d$ qubits. Since the area between the two squares is simply $w^2 - (w-2\ell)^2 = O(w\cdot \ell)$, the boundary contains less than $d$ qubits as long as $w \le O(d/\ell)$. Thus, we can ``grow'' a correctable square of size up to $w = \Theta(d/\ell)$. An illustration of the ``growing-the-square'' process is given in Figure \ref{fig:growing-the-square}.

\begin{figure}
    \centering
      \begin{tikzpicture}[scale=0.45]
    \draw[white,pattern=north west lines,pattern color=blue!40] (0,0) rectangle (15,15);
    \foreach \i in {0,5,...,15} {
      \draw [pattern=crosshatch,pattern color=red!100] (\i-0.2,0-0.2) rectangle (\i+0.2,15+0.2);
      \draw [pattern=crosshatch,pattern color=red!100] (0-0.2,\i-0.2) rectangle (15+0.2,\i+0.2);
    }
    \foreach \i in {0,5,...,15} {
      \foreach \j in {0,5,...,15} {
        \draw[fill=yellow!90] (\i-0.4,\j-0.4) rectangle (\i+0.4,\j+0.4);
      }
    }
    
    \node (a) at (1,17) {$A$};
    \draw[->] (a) to[out=300,in=90] (2.5,12.5);

    \node (b) at (7.5, 17) {$B$};
    \draw[->] (b) to (7.5, 15);

    \node (c) at (14, 17) {$C$};
    \draw[->] (c) to (15,15);

    \node (l1) at (9, 7.5) [circle,fill,inner sep=1.5pt]{}; 
    \node (l2) at (11, 7.5) [circle,fill,inner sep=1.5pt]{};
    \draw[thick] (l1) -- (l2);
  \end{tikzpicture} 
  \caption{The partition into sets $A, B, C$ when there are (1) no long interactions and (2) qubits are arranged in a lattice. $A$ is the union of the blue squares, $B$ is the union of the red rectangles, and $C$ is the union of the yellow squares. A long interaction is drawn connecting two previously disconnected blue squares.}
  \label{fig:proof-sketch-partition}
\end{figure}
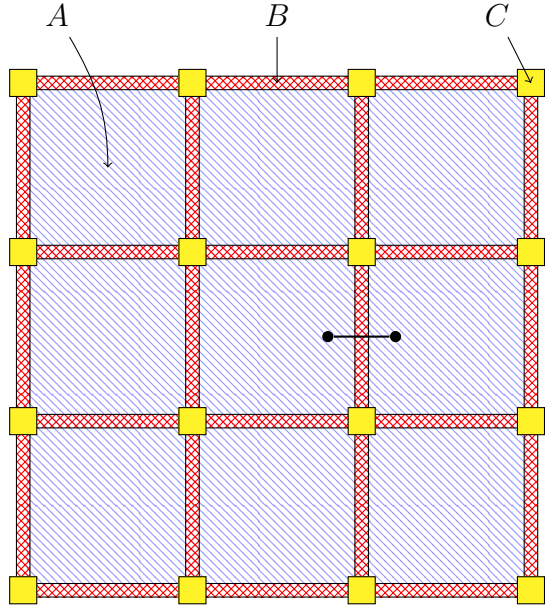

We now partition our qubits into sets $A,B,C$ by partioning our $\sqrt{n}\times \sqrt{n}$ grid (see Figure~\ref{fig:proof-sketch-partition}). The set $A$ consists of $\Theta(d/\ell)\times \Theta(d/\ell)$ squares in the plane, separated by $\Theta(d/\ell) \times \Theta(\ell)$ rectangular edges, which form the set $B$. We remove a $\Theta(\ell) \times \Theta(\ell)$ square from every corner of this grid, and place these qubits in $C$. The sets $A,B,C$ are constructed so that each edge in $B$ separates two horizontally or vertically neighboring squares in $A$, acting as a barrier of size at least $\ell$. Similarly, the corner squares in $C$ separate diagonally neighboring squares in $A$ and rectangles in $B$.

We now check that this our desired partition. Note that $A$ is a union of correctable squares grown from the method described above. Further these squares are decoupled because they are at pairwise distance at least $\ell$ and we assume no interactions of length at least $\ell$, so by the Union Lemma (Lemma~\ref{lem:correctable}), $A$ is correctable. Similarly, each rectangle in $B$ has size $O(d/\ell)\times O(\ell) < d$, so they are correctable as well. The rectangles in $B$ are also at pairwise distance at least $\ell$, so the rectangles are pairwise decoupled, and $B$ is thus correctable itself by the Union Lemma. Lastly, we bound the size of $C$. Note that each corner square contains $O(\ell^2)$ qubits, and the number of squares is simply $\frac{\sqrt{n}}{\Theta(d/\ell)} \times \frac{\sqrt{n}}{\Theta(d/\ell)} = O(\frac{n\ell^2}{d^2})$. Thus, $C$ contains $O(\ell)^2 \cdot O(\frac{n\ell^2}{d^2}) = O(\frac{n\ell^4}{d^2})$ qubits, and we conclude by Lemma~\ref{lem:bptabc} that $k\le |C|\le O(\frac{n\ell^4}{d^2})$, as desired.

\subsection{Long-range interactions}

We now discuss how to prove a better bound on the number of interactions.
Because there are now more parameters, we change the proof structure slightly by using contradiction.
The overall structure remains the same: we assume there are few long interactions, and then construct a partition $A \sqcup B \sqcup C$ so that $A$ and $B$ are correctable and $C$ is small.
However, we now explicitly quantify what ``few'' and ''long'' both mean --- to be concrete, we assume our code has at most $\frac{\max(k,d)}{100}$ interactions of length at least $\ell\defeq \frac{1}{100}\big(\frac{kd^2}{n}\big)^{1/4}$ --- and then show that $\abs{C} < k$, contradicting Lemma \ref{lem:bptabc}. We conclude that our code must have more long interactions than was prescribed. We note that the above proof of Proposition~\ref{pr:main} can also be presented in this format.

Beyond this structural change, we develop new strategies to handle the long interactions that appear in our embeddings. 
The first potential problem with long interactions is that we cannot directly apply the simple method to grow large correctable sets from smaller ones, since now the large outer square may not contain all neighbors of qubits within the smaller internal square (see Figure \ref{fig:growing-the-square}). Fortunately, this is not an issue if each large square participates in at most $d/10$ long interactions (see Lemma~\ref{lem:square-1}); our application of the Expansion Lemma only requires that the size of the boundary be less than $d$. 
Thus, we can still apply the Expansion Lemma if there are at most $9d/10$ boundary qubits from short interactions,
and we can grow a correctable square up to size $O(d/\ell)$ (with a slightly worse constant), just as before.
An important technical detail is that, after each application of the Expansion Lemma, we use the Subset Closure property of correctable sets to remove the spurious qubits added from the long interactions, so that the Expansion Lemma is always applied only to the qubits in our growing square.

The second potential problem with long interactions is that long interactions can connect squares in $A$ that were previously disconnected by $O(\ell)$-width rectangles in $B$, which would prevent us from concluding $A$ is correctable with the Union Lemma. We use two different solutions to handle this problem, depending on the relative values of the dimension $k$ and the distance $d$, since we allow for $\frac{\max(k,d)}{100}$ long interactions.

When $k \geq d$ (see Figure~\ref{fig:abc} for illustration of partition), some squares in $A$ can contain more than the $d/10$ allowed long interactions. To deal with this problem, we subdivide these squares into smaller correctable rectangles (Lemma~\ref{lem:square-3}). Each subdivision adds a constant number of $O(\ell)\times O(\ell)$ squares to $C$, and we can show that an optimal $O(k/d)$ subdivisions suffice, for a total of $O(\ell^2\cdot k/d) < O(k)$ additional qubits to $C$. Further, we add the qubits of the $\frac{k}{100}$ long interactions to $C$, so that the total number of qubits in $C$ is still less than $k$. Since all qubits in $A$ and $B$ only use short interactions, we can thus conclude $A$ and $B$ are each correctable by the Union Lemma, as before.

When $d \geq k$ (see Figure~\ref{fig:abc} for illustration of partition), the set of all qubits participating in long interactions is correctable because there are less than $d$ such qubits (Distance Property, Lemma~\ref{lem:correctable}). Moreover, by a probabilistic averaging argument (Lemma~\ref{lem:square-2}), we can ensure most of these long-interaction qubits do not interact with qubits in $B$ (the squares in $A$ are much larger by area than the rectangles in $B$), so we can add most of these to $B$ while maintaining correctability. We add the few remaining long-interaction qubits to $C$, and can check we add at most $k$ qubits; the averaging technique ensures $O(\frac{\ell}{d/\ell})$ of the $O(d)$ long-interaction qubits end up in $B$, so there are at most\footnote{This bound assumes $d\le \sqrt{kn}$. We may assume this, as the case $d\ge \sqrt{kn}$ is covered by Theorem~\ref{thm:main-2}.} $O(\ell^2)\le O(k)$ such qubits added to $C$.  

\subsection{Moving off the lattice}
The second generalization is when we no longer assume that the qubits are arranged in a 2D-lattice. Previously, we used the grid assumption to show that a $s \times s$ square contains $s^2$ qubits. The key point here is that the number of qubits that can be contained in a square of area $A$ is $O(A)$.  
We show in \ref{lem:square-0} that this also holds when the qubits are not labeled on a 2D lattice. Intuitively, this is to be expected since qubits in a 2D embedding must be at least unit distance part, so most arbitrary embeddings will not effectively ``fill up'' the space with qubits. 

The other issue is maintaining the ``back of the envelope'' estimates of the number of qubits participating in long interactions in sets $A$ and $B$. This is because we need to re-associate the qubits participating in long interactions within our constructed sets $A,B$, and we need to know that there aren't too many bad interactions in a certain area to be ``absorbed" by another set.  
Lemma \ref{lem:square-2} shows that this is indeed the case: for any 2D-embedding of the qubits, we can find a grid tiling of the plane so that there is not too large of a concentration of qubits participating in long interactions in any particular area. 

\section{Preliminaries}
\label{sec:prelims}

\paragraph{Notation and Definitions.}

We use standard Landau notation $O(\cdot), \Omega(\cdot), \Theta(\cdot), o(\cdot), \omega(\cdot)$.
We also use the notations $\tilde O(\cdot), \tilde \Omega(\cdot)$, which are variants of $O(\cdot)$ and $\Omega(\cdot)$, respectively, that ignore log factors. For example, $f(n) =\tilde O(h(n))$ means that there exists an integer $k$ such that $f(n) = O(h(n) \log^k n)$.

\emph{The plane} refers to $\mathbb{R}^2$. 
In the plane, \emph{distance} refers to Euclidean ($\ell_2$) distance unless otherwise specified. 
We sometimes use the \emph{$\ell_\infty$-distance} of two points $(x,y),(x',y')\in\mathbb{R}^2$, which, recall, is $\max(|x-x'|,|y-y'|)$.
A \emph{grid tiling} is a division of the plane given by vertical and horizontal lines equally spaced at a fixed distance $w$.
Throughout, a \emph{rectangle} is always a set of the form $[a,b]\times[c,d]$. In particular, rectangles contain their boundary and are axis-parallel.
The same is true for squares.

A \emph{2D-embedded} set is a finite set $Q\subset\mathbb{R}^2$ with pairwise ($\ell_2$) distance at least 1.
A function $f:\mathbb{R}^2\to\mathbb{N}$ is \emph{finitely supported} if $f(x)\neq 0$ for finitely many $x\in\mathbb{R}^2$.
  For a finitely supported function $f:\mathbb{R}^2\to\mathbb{N}$ and region $R\subset\mathbb{R}^2$, define, by abuse of notation, $f(R)=\sum_{i\in R; f(i)\neq 0} f(i)$. 
We study the finitely supported function given by Definition~\ref{def:f}.

\subsection{Quantum codes}

\paragraph{Stabilizer Codes.}

We associate the pure states of a qubit with $\bbC^{2}$ and pure $n$ qubit pure states with $(\bbC^{2})^{\otimes n}$.
Let $\mathcal{P}$ denote the Pauli group, which consists of the Pauli set $\ssI, \ssX, \ssY, \ssZ$, and their scalar multiples by $\{\pm 1,\pm i\}$.
A stabilizer code $\mathcal{Q} = \mathcal{Q}(\mathcal{S})$ is specified by the stabilizer group $\mathcal{S}$, an abelian subgroup of the $n$-qubit Pauli group $\mathcal{P}_n = \mathcal{P}^{\otimes n}$ that does not contain $-\ssI$.
The code $\mathcal{Q}$ is the set of states left invariant under the action of the stabilizer group, i.e.\ $ \mathcal{Q} = \{ \ket{\psi} : \; \ssS \ket{\psi} = \ket{\psi} \;\forall \ssS \in \mathcal{S}\}$.
Being an abelian group, we can describe $\mathcal{S}$ by $n-k$ independent generators $\{\ssS_1,...,\ssS_{n-k}\}$, where $k$ is the \emph{dimension} of the code. The \emph{distance} $d$ is the minimum weight of an error $E$ that maps a codeword in $\mathcal{Q}$ to another codeword. A quantum code $\mathcal{Q}$ with distance $d$ can correct up to $d-1$ qubit erasures.

\paragraph{Quantum codes in 2D.}
  Throughout, we identify the qubits $Q$ of our quantum code with a 2D-embedded set, which we call $Q\subset \mathbb{R}^2$ by abuse of notation. By further abuse of notation, we refer to $Q\subset\mathbb{R}^2$ as the \emph{embedding} of the qubits $Q$.

  When the set of qubits $Q$ is understood, for a qubit subset $V\subset Q$, we let $\overline{V}\defeq Q\setminus V$ denote the \emph{complement} of $V$.

\begin{definition}[Interactions]
  Given a 2D-embedding $Q\subset \mathbb{R}^2$ of a quantum stabilizer code with a set of generators $\{\ssS_1,\dots,\ssS_{n-k}\}$ of its stabilizer group, an \emph{interaction} is a pair of qubits $p,q\in Q$ such that $p$ and $q$ are both in the support of some generator $\ssS_i$.
    The \emph{length} of the interaction is the distance between $p$ and $q$.
    An interaction is always defined with respect to a particular set of stabilizer generators, but throughout we assume that the generator set is fixed, and thus the set of interactions is fixed as well.
\end{definition}

\begin{remark}[Connectivity Graph]
    A helpful intuition for our argument, which past works have leveraged explicitly \cite{baspin2022connectivity,baspin2022quantifying,baspin2023improved,baspin2023lower}, is that the interactions of a quantum code form a graph, called the \emph{connectivity graph}, whose vertices are the qubits and whose edges are the interactions (with respect to a fixed generator set).
    We do not use this graph theoretic language explicitly as in prior works, though this connection motivates some of our proofs.
    \label{rem:graph}
\end{remark}

In each proof, we care about one particular interaction length $\ell$.
When $\ell$ is implicit, we informally refer to an interaction as \emph{bad} if the length is at least $\ell$ and \emph{good} if the interaction has length at most $\ell$ (intuitively, good interactions are easier to deal with for our proof).

We use the following function that counts the number of bad interactions that a particular qubit participates in.
\begin{definition}[Interaction counter]
  Given a quantum code with a 2D-embedding $Q\subset \mathbb{R}^2$, let $f_{\ge\ell}:\mathbb{R}^2\to \mathbb{N}$ denote the \emph{interaction counting function}, where $f_{\ge\ell}(q)$ for $q\in Q$ equals the number of interactions of length at least $\ell$ that qubit $q$ participates in, and $f_{\ge\ell}(\cdot)=0$ outside of $Q$.
  \label{def:f}
\end{definition}

\paragraph{Correctable sets.} Like in previous works \cite{baspin2022connectivity,baspin2022quantifying,baspin2023improved}, we analyze the limitations of quantum codes using correctable sets. Intuitively, a set $U$ of qubits is correctable with respect to a code if the code can correct the erasure of the qubits in $U$.
We state the definition of a correctable set for completeness, though we only interface with the definition indirectly, using Lemma~\ref{lem:correctable} and Lemma~\ref{lem:bptabc}.

\begin{definition}[Correctable set]
  For a set of qubits $U \subset Q$ in a quantum code, let $\mathcal{D}[\overline{U}]$ and $\mathcal{D}[Q]$ denote the space of density operators associated with the sets of qubits $\overline{U}$ and $Q$ respectively.
  The set $U$ is \emph{correctable} if there exists a recovery map $\mathcal{R}: \mathcal{D}[\overline{U}] \to \mathcal{D}[Q]$ such that for any code state $\rho$, we have $\mathcal{R}(\Tr_{U}(\rho)) = \rho$.
\end{definition}

We use the following notions in Lemma~\ref{lem:correctable} to reason about correctable sets.
In a quantum code with qubits $Q$, say sets $U_1,\dots,U_\ell\subset Q$ are \emph{decoupled} if there are no interactions between two distinct $U_i$'s.
For a set $U\subset Q$, let $\partial_+ U$ be the \emph{boundary} of $U$, the set of qubits outside $U$ that have an interaction with $U$.
\begin{lemma}[\cite{bravyi2009no,flammia2017limits}, see also \cite{baspin2023improved}]
  Let $Q$ be the set of qubits of a $[[n,k,d]]$ code.
\label{lem:correctable}
\begin{enumerate}
   \item \label{lem:trivial} \textbf{\emph{Subset Closure:}} Let $U \subseteq Q$ be a correctable set. 
   Any subset $W \subseteq U$ is correctable.
   \item \label{lem:dist} \textbf{\emph{Distance Property:}} If $U \subseteq Q$ such that $|U| < d$, then $U$ is correctable.
   \item \label{lem:bptunion} \textbf{\emph{Union Lemma:}}
    If $U_1,\dots,U_\ell$ are decoupled and each $U_i$ is correctable, then $\bigcup_{i=1}^\ell U_i$ is correctable.
    \item \label{lem:bptexpansion} \textbf{\emph{Expansion Lemma:}}
    If $U,T\subseteq Q$ are correctable sets such that $T \supseteq \partial_+ U$, then $T \cup U$ is correctable. 
\end{enumerate}
\end{lemma}

\begin{lemma}[Bravyi-Poulin-Terhal \cite{bravyi2010tradeoffs}, Eq.~14]
	\label{lem:bptabc}
	Suppose that the qubits $Q$ of an $[[n,k,d]]$ stabilizer code can be partitioned as $Q = A \sqcup B \sqcup C$ where $A$ and $B$ are correctable. Then,
 	\begin{equation*}
		k  \leq |C|~.
	\end{equation*}
\end{lemma}

The structure of all of our lower bounds is as follows: assume for sake of contradiction that our code has a lot of locality.
We find, with the help of Lemma~\ref{lem:correctable}, correctable sets $A$ and $B$ large enough that $n-|A|-|B| < k$.
In this case, applying Lemma~\ref{lem:bptabc} with $C = Q-(A\cup B)$ gives $k\le |C|=n-|A|-|B| < k$, which is a contradiction.
We conclude that our code has the desired amount of non-locality.

\subsection{Geometric Lemmas}

We establish two lemmas about 2D-embedded sets. 
These lemmas allow us extend beyond a $\sqrt{n}\times \sqrt{n}$ lattice embedding to handle arbitrary 2D-embeddings.
Further, Lemma~\ref{lem:square-2} uses an averaging argument to ensure that the distribution of long-range interactions in our embedding lines up with back-of-the-envelope estimates.
\begin{lemma}
  Suppose $Q\subset \mathbb{R}^2$ is 2D-embedded and $R$ is a rectangle with area $A$. (i) If $R$ has both side lengths at least 1, then at most $6A$ points of $Q$ lie in $R$, i.e., $|Q\cap R|\le 6A$.
  (ii) If $R$ has one side length $s$ and another side length less than 1, then $|Q\cap R|\le 4s$.
  \label{lem:square-0}
\end{lemma}
\begin{proof} 
  Proof by pigeonhole.
  First suppose $R$ has side lengths least 1.
  Draw a circle of radius 1/2 around each point. Each circle has at least $\frac{\pi}{4}\cdot (1/2)^2\ge 0.19$ units of area inside $R$, and all circles are disjoint since $Q$ is 2D-embedded. Then $0.19\cdot|Q\cap R| \le A$, so $|Q\cap R| < 6A$.

  Now suppose $R$ is $s\times t$ where $t<1$.
  Draw a circle of radius 1/2 around each point. Each circle has at least $t/4$ units of area inside $R$ (the circle-intersect-rectangle contains two triangles whose heights sum to $t$ and whose base is 1/2, for a total area of $t/4$), and all circles are disjoint since $Q$ is 2D-embedded. Then $(t/4)\cdot|Q\cap R| \le A=t\cdot s$, so $|Q\cap R| < 4s$.
\end{proof}

\begin{figure}
    \centering
    \begin{tikzpicture}[scale=1]
        \draw[draw=none,fill=orange!60] (-2.,-2.) rectangle (2,2);
        \draw[draw=none,fill=yellow!60] (-1.5, -1.5) rectangle (1.5, 1.5);
        \draw[draw=none,fill=red!60] (-2,-2) rectangle (-1.5,-1.5);
        \draw[draw=none,fill=red!60] (-2,2) rectangle (-1.5,1.5);
        \draw[draw=none,fill=red!60] (2,-2) rectangle (1.5,-1.5);
        \draw[draw=none,fill=red!60] (2,2) rectangle (1.5,1.5);
        \draw[line width=2pt] (-2,-4) -- (-2, 4);
        \draw[line width=2pt] (2,-4) -- (2, 4);
        \draw[line width=2pt] (-4,-2) -- (4, -2);
        \draw[line width=2pt] (-4,2) -- (4,2);
        \node[rotate=45] at (3,3) {\Large $\cdots$};
        \node[rotate=45] at (-3,-3) {\Large $\cdots$};
        \node[rotate=135] at (-3,3) {\Large $\cdots$};
        \node[rotate=135] at (3,-3) {\Large $\cdots$};

        \draw[<->, dotted, thick] (-2, -2.15) -- (2, -2.15) node [midway, below] {$w$};
        \draw[<->, dotted, thick] (-2., -1) -- (-1.5, -1) node [midway, above] {$2\ell$};

        \node[] (corner-label) at (6, 0) {$O(\ell^2/w^2)$ fraction here};
        \node[] (side-label) at (-6,0) {$O(\ell/w)$ fraction here};
    \draw[->] (corner-label) to[out=140,in=0] (1.75,1.75);
    \draw[->] (side-label) to[out=330,in=120] (-1.75,0);
    \end{tikzpicture}
\caption{Tiling Lemma: for fixed sets of points $X$ and $Y$ and a random width-$w$ grid tiling, we expected $O(\ell^2/w^2)$ fraction of $X$ to be within a $O(\ell)$ of a grid vertex, and $O(\ell/w)$ fraction of $Y$ to be within $O(\ell)$ of an grid edge.}
\label{fig:tiling}
\end{figure}
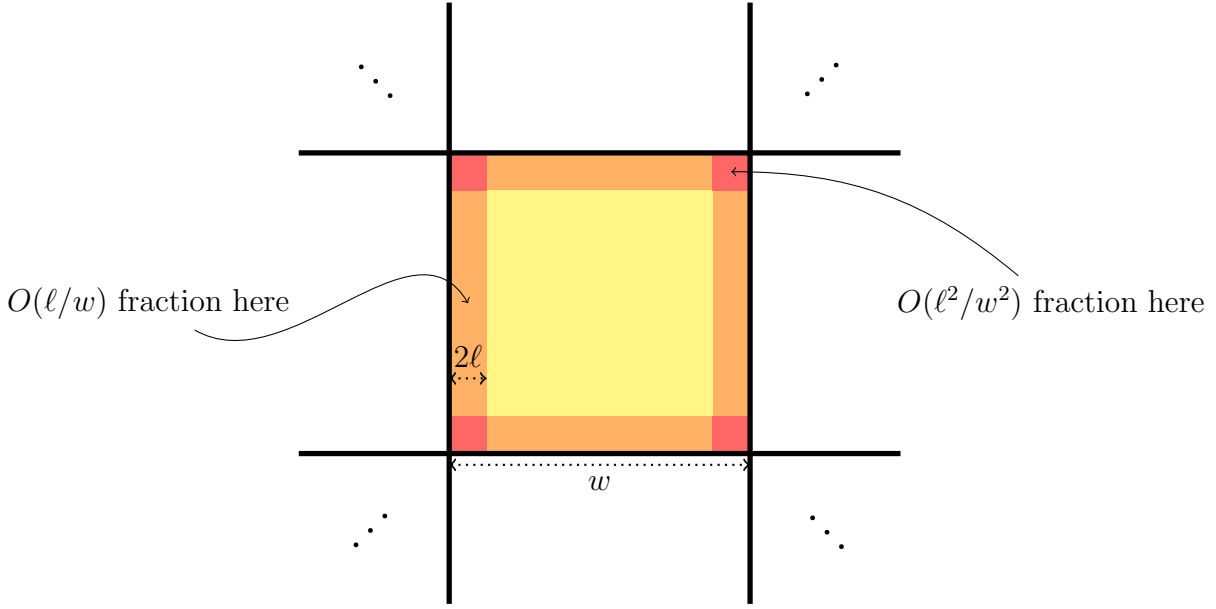

We use the next lemma (see also, Figure~\ref{fig:tiling}) --- an application of the probabilistic method \cite{alon2016probabilistic} --- in the main theorem, Theorem~\ref{thm:main}, to generate the grid tiling that produces the partition $A\sqcup B\sqcup C$ in our proof.
The technical guarantees of Lemma~\ref{lem:square-2} help generalize our main theorem from stabilizer codes with the standard lattice embedding to stabilizer codes with arbitrary 2D embeddings.
The guarantees on $Y$ also ensure that the distribution of long-range interactions in our embedding lines up with back-of-the-envelope estimates.

\begin{lemma}[Tiling Lemma]
  Let $X,Y\subset\mathbb{R}^2$ be two multisets.
  Let $w$ and $\ell$ be positive integers with $w\ge 4\ell$.
  There exists a grid tiling of $\mathbb{R}^2$ by $w\times w$ squares such that (i) at most $\frac{32\ell^2}{w^2}$ fraction of the points in $X$ are within $\ell_\infty$-distance $2\ell$ of the vertices of some square and (ii) at most $\frac{16\ell}{w}$ fraction of the points in $Y$ are within distance $2\ell$ of any edge of any square.
  \label{lem:square-2}
\end{lemma}
\begin{proof}
  The proof is by the probabilistic method.
    Let $x$ and $y$ be independently sampled uniformly from the interval $[0,w]$. 
    Consider the grid tiling of $\mathbb{R}^2$ by $w\times w$ squares with origin at $(x,y)$. 
    Say a point in $X$ is \emph{$X$-bad} if it is within $\ell_\infty$ distance $2\ell$ of some vertex of some square.
    Say a point in $Y$ is \emph{$Y$-bad} if it is within distance $2\ell$ of some edge of some square.
    
    The probability that a point $a\in X$ is $X$-bad is exactly $\frac{16\ell^2}{w^2}$.
    The probability that a point $b\in Y$ is $Y$-bad is $\frac{w^2-(w-4\ell)^2}{w^2}< \frac{8\ell}{w}$.
    Thus, the expected number of $X$-bad points is at most $\frac{16\ell^2}{w^2}\cdot |X|$, and the expected number of $Y$-bad points is at most $\frac{8\ell}{w}\cdot |Y|$.
    By Markov's inequality, the probability the number of $X$-bad points is more than than $\frac{32\ell^2}{w^2}\cdot |X|$ is less than 1/2.
    Similarly, the probability the number of $Y$-bad points is more than $\frac{16\ell}{w}\cdot |Y|$ is less than 1/2.
    Thus, there exists some choice of $(x,y)$, and thus some grid tiling, where there are at most $\frac{32\ell^2}{w^2}\cdot |X|$ $X$-bad points and at most $\frac{16\ell}{w}\cdot |Y|$ $Y$-bad points.
\end{proof}

\section{Proof of Main Theorem}
\label{sec:proof}

We now prove Theorem~\ref{thm:main-2} and Theorem~\ref{thm:main-3}, which together imply our main theorem, Theorem~\ref{thm:main}.

\subsection{Proof of Theorem~\ref{thm:main-2}}
Our proof of Theorem~\ref{thm:main-2} is inspired by \cite[Lemma 3.5]{baspin2023improved}. We first show (Lemma~\ref{lem:plane-1}) that any subset of qubits has a good ``separator", a set of points whose removal divides the graph into two disjoint and almost-disconnected parts.
To prove Theorem~\ref{thm:main-2}, under our contradiction assumption, we recursively divide the vertex set in half using separators, until all the sets have size less than $d$ and are correctable. 
We then alternately apply the Union and Expansion Lemmas to deduce that larger and larger qubit sets are correctable, until we show the set of all qubits is correctable, giving our contradiction.

This proof imitates \cite[Lemma 3.5]{baspin2023improved}, but with one key difference.
In \cite[Lemma 3.5]{baspin2023improved}, the separator divides a vertex subset into two disjoint and disconnected parts, our separator only divides it into disjoint and \emph{almost-}disconnected parts; the parts are not entirely disconnected due to the presence of bad interactions. However, because the number of bad interactions is small, we can always work around them when we apply the Union and Expansion lemmas.

We now state and prove our separator lemma, Lemma~\ref{lem:plane-1}.
\begin{lemma}
    Let $\ell$ be a positive real number and $n\ge 10$ be a positive integer.
    Give a set $V$ of $n$ 2D-embedded points, there exists two parallel lines that divide the plane into three regions such that, (i) the lines are at distance at least $\ell$, (ii) the region between the two lines has at most $8\ell\sqrt{n}$ points, and (iii) the other two regions each have at most $9n/10$ points.
    Further, we can ensure (iv) the two lines are either both horizontal or both vertical, and no point lies on any line. 
    \label{lem:plane-1}
\end{lemma}
\begin{figure}
  \centering
  \begin{tikzpicture}
    \draw[white,fill=red!10] (-4.2,0) rectangle (4.2,0.5);
    \node[red!30,anchor=west] at (4.2,0.25) {$\le 8\ell\sqrt{n}$ qubits};
    \draw[thick,<->] (-2,-3) -- (-2,3) node [at end,above] {$y=a_1$};
    \draw[thick,<->] (2,-3) -- (2,3) node [at end,above] {$y=a_2$};
    \draw[thick,<->] (-3,-2) -- (3,-2) node [at end,right] {$y=b_1$};
    \draw[thick,<->] (-3,2) -- (3,2) node [at end,right] {$y=b_2$};
    \node[gray] at (0,2.8) {$\ge n/10$};
    \node[gray] at (0,-2.8) {$\ge n/10$};
    \node[gray] at (0,0) {$\ge 6n/10$};
    
    \foreach \y in {-1.5,-1,...,1.5} {
      \draw[<->] (-4,\y) -- (4,\y);
    }
    \draw[<->,gray] (-3.2,1) -- (-3.2,1.5) node [midway, left] {$\ell$};
    \draw[<->,gray] (-4.4,-2) -- (-4.4,2) node [midway, left] {$\ge \sqrt{n/10}$};
  \end{tikzpicture}
  \caption{Lemma~\ref{lem:plane-1}, our separator lemma. Any set of $n$ 2D-embedded points has a width-$\ell$ separator with at most $8\ell\sqrt{n}$ points. The red region is our separator.}
  \label{fig:separator}
\end{figure}
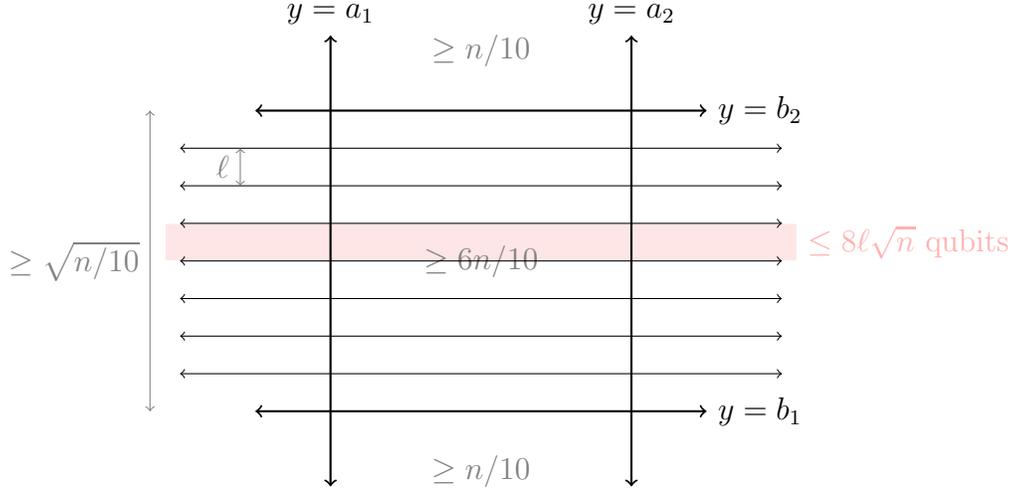

\begin{proof}
    We may assume $\ell\le \sqrt{n}/8$ or else the statement is trivial.
    Define
    \begin{itemize}
        \item $a_1$ is the minimum value such that there are at least $n/10$ points $(x,y)\in V$ with $x\le a_1$
        \item $a_2$ is the maximum value such that there are at least $n/10$ points $(x,y)\in V$ with $x\ge a_2$
        \item $b_1$ is the minimum value such that there are at least $n/10$ points $(x,y)\in V$ with $y\le b_1$
        \item $b_2$ is the maximum value such that there are at least $n/10$ points $(x,y)\in V$ with $y\ge b_2$
    \end{itemize}
    We claim that either $b_2-b_1 > \sqrt{n/10}$ or $a_2-a_1 > \sqrt{n/10}$.
See Figure~\ref{fig:separator} for an illustration of the division.

    By the extremality of $a_1,a_2,b_1,b_2$, for each of (i) $x<a_1$, (ii) $x>a_2$, (iii) $y<b_1$, (iv) $y>b_2$, there are less than $n/10$ points $(x,y)\in V$ with that property. Thus, there are at least $n-4\cdot (n/10) > 6n/10$ points inside the rectangle $R\defeq [a_1,a_2]\times [b_1,b_2]$. 
    By Lemma~\ref{lem:square-0}, one side length is at least $\sqrt{n/10}$: either the area of this rectangle must be at least $n/10$, or one side length is at least $\frac{1}{4}(6n/10) > \sqrt{n/10}$.
    Hence, we must either have $b_2-b_1 > \sqrt{n/10}$ or $a_2-a_1>\sqrt{n/10}$.
    Without loss of generality, $a_2-a_1>\sqrt{n/10}$.
    Then, we can divide the rectangle $R$ into $\floor{\frac{\sqrt{n/10}}{\ell}} > \frac{\sqrt{n}}{8\ell}$  parts, using a collection of vertical lines at pairwise distance greater than $\ell$ and that include the lines $x=a_1$ and $x_2=a_2$. 
    By pigeonhole, one of these parts, between the lines $x=a_1', x=a_2'$ has at most $ 8\ell\sqrt{n}$  points of $V$.  
    These lines $x=a_1', x=a_2'$ satisfy (i) and (ii) by definition, and (iii) because $a_1\le a_1'\le a_2'\le a_2$.
    We can further guarantee (iv) by slightly perturbing $a_1'$ and $a_2'$.
\end{proof}

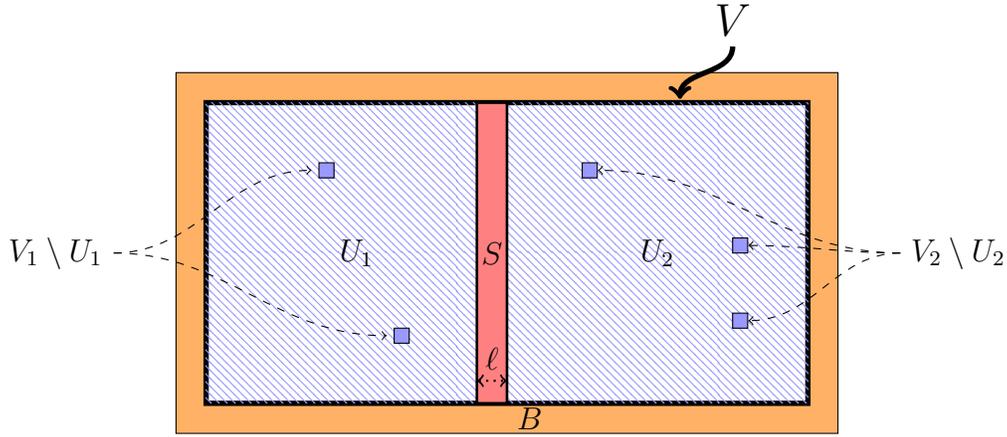
\begin{figure}
    \centering
    \begin{tikzpicture}[scale=1]
        \draw[fill=orange!60] (-4.4,-2.4) rectangle (4.4,2.4);
                \draw[line width=2pt,fill=white] (-4,-2) rectangle (4,2);
        \draw[draw=none,fill=white,pattern=north west lines, pattern color=blue!40] (-4.,-2.) rectangle (4,2);
        \draw[line width=1pt,fill=red!50] (-0.4,-2) rectangle (0,2);
        \node[] () at (-0.2,0) {$S$};

         \draw[<->, dotted, thick] (-0.4, -1.7) -- (0, -1.7) node [midway, above] {$\ell$};

        \node[] (v) at (3,3.1) {\Large $V$};
        \node[] (b) at (0.3,-2.2) {$B$};

         \node[] () at (-2,0) {$U_1$};
         \node[] () at (2,0) {$U_2$};

         \node[] (v1) at (-6,0) {$V_1\setminus U_1$};
         \node[] (v2) at (6,0) {$V_2\setminus U_2$};
         \foreach \x / \y in {-2.5/1,-1.5/-1.2} {
          \draw[fill=blue!40] (\x,\y) rectangle +(.2,.2);     
           \draw[->,dashed] (v1) to[out=0,in=180] (\x-0.1,\y+0.1);
         }
         \foreach \x / \y in {1/1,3/0,3/-1} {
            \draw[fill=blue!40] (\x,\y) rectangle +(.2,.2);    
            \draw[->,dashed] (v2) to[out=180,in=0] (\x+0.21,\y+0.1);
         }
         \draw[->,line width=2pt] (v) to[out=270,in=90] (2.3,2.05);
    \end{tikzpicture}
\caption{Claim~\ref{cl:1}. Show a set is correctable by repeatedly subdividing using separators. Here we subdivide a set $V\subset Q$ using a separator $S$ (solid red) into two parts $V_1$ and $V_2$ (lined and solid blue). We then remove the bad qubits (solid blue) to get $U_1$ and $U_2$ (lined blue). By induction, $V_1$ and $V_2$ are correctable, so $U_1$ and $U_2$ are correctable (Subset Closure). By the Union Lemma $U_1\cup U_2$ is correctable. The boundary of $U_1\cup U_2$ is contained in the union of  $S$, $V_1\setminus U_1$, $V_2\setminus U_2$, and $B$ (the good qubits in the boundary of $V$) all of which are small and thus together correctable by the Distance Property. Hence, by the Expansion Lemma, $V$ is correctable.}
\label{fig:cl}
\end{figure}

We now prove Theorem~\ref{thm:main-2}.
\begin{proof}[Proof of Theorem~\ref{thm:main-2}]
  With hindsight, choose $c_0=\frac{1}{200}$ and $c_1=100$.
  Assume for sake of contradiction that a code with $kd^2\ge 100n$ has at most $d/200 < d/10$ interactions of length at least $\ell\defeq \frac{d}{200\sqrt{n}}$.
  We first handle the corner case of $\ell < 1$.
  Here, there are  at most $d/10$ interactions of any length (each interaction length is at least 1 by definition of embedding). By the Distance Property (Lemma~\ref{lem:correctable}), the set of at-most-$d/5$ qubits in an interaction are correctable, and any single qubit is correctable (since $kd^2 \ge 100n$, then $d\ge 10>1$). 
  By the Union Lemma (Lemma~\ref{lem:correctable}) the set of all qubits is correctable, and we have $k=0$, obtaining our contradiction.
  Thus, for the rest of the proof, assume $\ell\ge 1$.
  We prove by induction the following technical claim.
  \begin{claim}
      Let $V\subset Q$ be the subset of qubits lying in some rectangle $R$. Suppose that $f_{\ge\ell}(V)\le d/10$ and the number qubits in $\overline{V}$ that participate in good (length $\le\ell$) interactions with qubits in  $V$ is at most
      \begin{align}
      8\ell\sqrt{n}\cdot\sum_{i=0}^{\floor{\log_{9/10}(|V|/n)}-1} \sqrt{\frac{9}{10}}^i
      \label{eq:cl}
      \end{align}
  Then $V$ is correctable. (for intuition, note that \eqref{eq:cl} is at most $8\ell\sqrt{n} \cdot \frac{1-\sqrt{|V|/n}}{1-\sqrt{9/10}}\le 160\ell\sqrt{n}$.) 
  \label{cl:1}
  \end{claim}
  \begin{proof}
  We prove by induction on the number of points in $V$. First, if $|V| < d$, then $V$ is correctable by the Distance Property (Lemma~\ref{lem:correctable}), so assume otherwise. In particular, $|V|\ge d\ge 10$.
  
  Now suppose $V$ is a set of qubits satisfies $f_{\ge\ell}(V)\le d/10$ and \eqref{eq:cl}, and assume for induction the Claim is true for all $|V'| < |V|$.
  Apply Lemma~\ref{lem:plane-1} to $V$ to obtain two parallel lines at distance at least $\ell$. 
  These partition $V$ into are three sets $V_1, V_2, S$, where $S$ is the qubits between the lines.  Because $V_1$ and $V_2$ each contain points of $V$, the two lines pass through the interior of $R$.
  We check the induction hypothesis applies to $V_1$: 
  \begin{itemize}
    \item  $|V_1| < \frac{9}{10}|V|$
    \item $f_{\ge\ell}(V_1) \le f_{\ge\ell}(V)\le d/10$.
    \item All qubits in $\overline{V_1}$ participating in good (length $\le \ell$) interactions with $V_1$ and must be either in $S$ or $\overline{V}$ (not $V_2$ because points in $V_2$ are at distance at least $\ell$ from $V_1$). By \eqref{eq:cl}, the number of such qubits is thus at most
    \begin{align}
      &\le |S|+ 8\ell\sqrt{n}\cdot\sum_{i=0}^{\floor{\log_{9/10}(|V|/n)}-1} \sqrt{\frac{9}{10}}^i\nonumber\\
      &\le 8\ell\sqrt{|V|}+ 8\ell\sqrt{n}\cdot\sum_{i=0}^{\floor{\log_{9/10}(|V|/n)}-1} \sqrt{\frac{9}{10}}^i \nonumber\\
      &<8\ell\sqrt{n}\cdot\sum_{i=0}^{\floor{\log_{9/10}(|V_1|/n)}-1} \sqrt{\frac{9}{10}}^i
    \end{align}
    In the last inequality, we used that $|V_1|\le \frac{9}{10}|V|$.
  \end{itemize}
  This shows that $V_1$ is correctable by the induction hypothesis. By a symmetric argument, $V_2$ is also correctable.
  Let $U_1\subset V_1$ and $U_2\subset V_2$ be the sets of qubits in $V_1$ and $V_2$, respectively, that only participate in good interactions.
  Since $V_1$ and $V_2$ are correctable, $U_1$ and $U_2$ are correctable by Subset Closure (Lemma~\ref{lem:correctable}).
  Since each qubit in $U_1$ and $U_2$ participates only in good interactions, and qubits in $U_1$ and $U_2$ are at distance at least $\ell$, set $U_1\cup U_2$ is correctable by the Union Lemma (Lemma~\ref{lem:correctable}).
  Lastly, let $B$ denote the qubits in $\overline{V}$ that participate in a good interaction with a qubit in $V$.
  Since $U_1$ and $U_2$ only contain good qubits, the set $T\defeq B\cup (V\setminus (U_1\cup U_2)$ contains the boundary of $U_1\cup U_2$. Further
  \begin{align}
      \abs{T}
      &\le |V_1\setminus U_1| + |V_2\setminus U_2| + |S| + |B| \nonumber \\
      &\le \frac{d}{10}+ 8\ell\sqrt{|V|} + 8\ell\sqrt{n}\cdot\sum_{i=0}^{\floor{\log_{9/10}(|V|/n)}-1} \sqrt{\frac{9}{10}}^i  \nonumber \\
      &\le \frac{d}{10} + 8\ell\sqrt{n}\cdot\sum_{i=0}^{\infty} \sqrt{\frac{9}{10}}^i \nonumber \\ 
      &= \frac{d}{10} + 8\ell\sqrt{n}\cdot\frac{1}{1-\sqrt{9/10}} \nonumber\\
      &< d
  \end{align}
  where the last inequality uses our choice of $\ell=\frac{d}{200\sqrt{n}}$.
  Thus, $T$ is correctable by the Distance Property (Lemma~\ref{lem:correctable}), so $T\cup (U_1\cup U_2)$ is correctable by the Expansion Lemma (Lemma~\ref{lem:correctable}).
  Hence, by Subset Closure (Lemma~\ref{lem:correctable}), $V\subset T\cup(U_1\cup U_2)$ is correctable.
  \end{proof}

We now apply Claim~\ref{cl:1} to a rectangle that contains all qubits $Q$. We have $f_{\ge\ell}(Q)\le d/10$ by assumption. Further, there are no qubits in $\overline{Q}$ so \eqref{eq:cl} holds trivially. Hence, the claim applies, and $Q$ is correctable. Since the set of all qubits is correctable, Lemma~\ref{lem:bptabc} implies that the dimension $k=0$, giving our contradiction.
\end{proof}

\subsection{Proof of Theorem~\ref{thm:main-3}: Two Lemmas}

Recall that, for a set of qubits $V$, the notation $f_{\ge\ell}(V)$ counts the number of times a qubit in $V$ participates in a bad interaction (length $\ge\ell$).
We first prove two lemmas that help us find correctable sets in a 2D-embedded quantum code.
The first lemma, Lemma~\ref{lem:square-1}, essentially follows as a corollary from Claim~\ref{cl:1}. 
However, in Lemma~\ref{lem:square-1}, we additionally assume the side lengths of the rectangle are bounded, resulting in a simpler proof. We present the simplified proof here for the reader's benefit (see also Section~\ref{sec:overview} and Figure~\ref{fig:growing-the-square}).
\begin{lemma}[Growing the square lemma]
  Let $d$ be a positive integer and $\ell\ge 1$.
  Suppose we have a $[[n,k,d]]$ quantum code with a 2D-embedding $Q\subset \mathbb{R}^2$.
  Let $V\subset Q$ be the set of all qubits lying in some rectangle with sides of length at most $\frac{d}{100\ell}$.
  If $f_{\ge\ell}(V)\le d/8$, then $V$ is correctable.
  \label{lem:square-1}
\end{lemma}
\begin{proof}
  The idea is to ``grow'' a correctable set from an initial $\sqrt{d}\times \sqrt{d}$ square, which we know is correctable by the Distance Lemma (Lemma~\ref{lem:correctable}).
  For $w\le \frac{d}{100\ell}$, an $(w-2\ell)\times (w-2\ell)$ correctable squares yields a $w\times w$ correctable square using the Expansion Lemma.

  By Subset Closure (Lemma~\ref{lem:correctable}), it suffices to prove the lemma for when $V$ is a square. 

We prove by induction on $w$, the side length of the square containing $V$.
For the base case, suppose $w\le \sqrt{d}/4$.
Then, $V$ lies in a square of area at most $d/16$, so $V$ has less than $d$ qubits by Lemma~\ref{lem:square-0}, so $V$ is correctable by the Distance Property (Lemma~\ref{lem:correctable}).

For the induction step, assume that $\alpha\ge \sqrt{d}/4$ and the statement is true for all squares of side length $\le \alpha$.
Now let $w$ be such that between $\alpha\le w\le \alpha+2\ell$ and $w<\frac{d}{100\ell}$.

Suppose the qubits of $V$ lie in a square $S$ of side length $s$ and let $U$ be the qubits inside the square $S'$ with the same center as $S$ and side length $w-2\ell$. By the induction hypothesis, $U$ is correctable. 
Let $T$ be the set of qubits that either (i) are in $S\setminus S'$ or (ii) participate in a bad (length $\ge \ell$) interaction with a qubit in $U$. 
By definition, $T\supset \partial_+ U$ ($T$ contains the boundary of $U$).
By Lemma~\ref{lem:square-0} (partition $S\setminus S'$ into four rectangles), there are at most $6(w^2 - (w-2\ell)^2)<24w\ell < d/4$ qubits of type (i), and by assumption there are at most $f_{\ge \ell}(U)\le f_{\ge \ell}(V)\le d/8$ qubits of type (ii).

Hence, $|T|<d$ and $T$ is correctable by the Distance Lemma (Lemma~\ref{lem:correctable}).
Thus, by the Expansion Lemma, $T \cup U$ is correctable, so $V$, a subset of $T\cup U$, is correctable as well by Subset Closure (Lemma~\ref{lem:correctable}).
This completes the induction, completing the proof.
\end{proof}

When we divide the plane into squares using Lemma~\ref{lem:square-2}, a back-of-the-envelope calculation says that the ``typical'' square is \emph{good}, meaning it has $\ll d$ bad interactions.\footnote{Omitting insignificant constants, the number of interactions is at most $\max(k,d)/100$. If $d\ge k$, this is at most $d/100$, so all squares are  good. Otherwise, the number of interactions is $O(k)$. 
There are $n$ qubits divided across squares that each contain $O(d^2/\ell^2)$ qubits, so the average number of interactions per square is approximately $O(\frac{kd^2}{n\ell^2})$. 
For $\ell=\sqrt[4]{\frac{kd^2}{n}}$, this is less than $d$, so the typical square is good.}
Lemma~\ref{lem:square-1} ensures that these good squares are correctable.
Lemma~\ref{lem:square-3} below handles the \emph{bad} squares, if any exist, by subdividing them into correctable rectangles (see Figure~\ref{fig:square-3}).
Since these bad squares are rare, the total cost incurred by these subdivisions is acceptable.

\begin{lemma}[Subdivision Lemma]
    Let $w,\ell$ and $d_1$ be positive real numbers.
    Let  $f:\mathbb{R}^2\to \mathbb{N}$ be a finitely supported function.
    Let $R$ be a $w\times h$ rectangle, with $h\ge 5\ell$ and $f(R)\ge d_1$ 
    Then there exists a division of $R$ by horizontal lines into rectangles $R_1,\dots,R_m$ such that 
    \begin{enumerate}
      \item Each rectangle has width $w$ and height $h_i$ with $h_i\ge 5\ell$.
      \item Each $R_i$ satisfies either (i) $f(R_i)\le d_1$ or (ii) has height $h_i\le 10\ell$. 
      \item The number of rectangles $m$ is at most $\frac{2f(R)}{d_1}$.
    \end{enumerate}
\label{lem:square-3}
\end{lemma}
  \begin{figure} 
  \begin{center}
  \begin{tikzpicture}
    \draw[fill=blue!20] (0,1)  rectangle (6,3);
    \node at (3,2) {$R[a,b]$};
    \draw[<->,] (4,0.04) -- (4,1-0.02) node [midway, left] {$a$};
    \draw[<->,] (5,0.04) -- (5,3-0.02) node [midway, right] {$b$};
    \draw[line width=2pt] (0,0)  rectangle (6,6);
  \end{tikzpicture}
  \qquad
  \begin{tikzpicture}
    \draw[line width=2pt, fill=blue!20] (0,0)  rectangle (6,6);
    \draw[line width=1pt, fill=blue!20] (0,0)  rectangle (6,1);
    \draw[line width=1pt, fill=blue!20] (0,1)  rectangle (6,4);
    \node at (3,0.5) {$R_1$};
    \node at (3,2.5) {$\cdots$};
    \node at (3,5) {$R_m$};
    \draw[line width=2pt] (0,0)  rectangle (6,6);
  \end{tikzpicture}
\end{center}
  \caption{Left: Subrectangle notation $R[a,b]$. Right: Lemma~\ref{lem:square-3}. Our division ensures that each region is correctable: either (i) it has few bad interactions, in which case is correctable by the Growing-the-Square-Lemma~\ref{lem:square-1}, or (ii) it has small area, in which case it is correctable by Lemma~\ref{lem:square-0} and the Distance Property.}
  \label{fig:square-3}
  \end{figure}
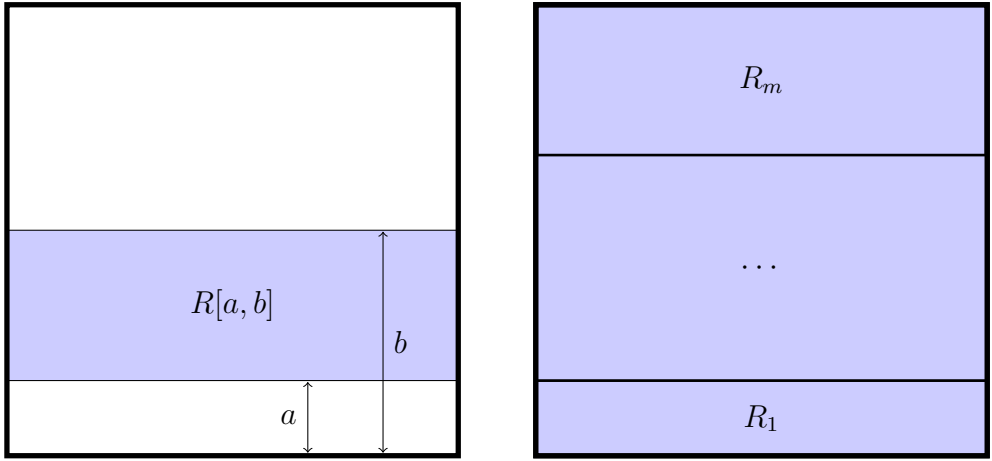
\begin{proof}
We prove by induction on $h$ and essentially take a greedy division.
If $5\ell\le h\le 20\ell$, then we take a division of $R$ into one or two rectangles of height between $5\ell$ and $10\ell$, and this satisfies the requirement.
The height(s) ensure properties 1 and 2 are satisfied, and the number of rectangles is at most $2\le 2\cdot \frac{f(R)}{d_1}$.

Now suppose $\alpha\ge 20\ell$ and assume for induction that the statement is true for all $h$ with $5\ell\le h\le \alpha$.
Now consider a $h$ with $\alpha\le h\le \alpha+5\ell$.
Suppose that $R$ is a $w\times h$ rectangle with $f(R) > d_1$.
For $0\le a\le b\le h$, let $R[a,b]$ denotes the $w\times (b-a)$ subrectangle of $R$ whose horizontal edges are distance $a$ and $b$ from the bottom edge of $R$ (see Figure~\ref{fig:square-3}).

Let $h_1'$ be the minimum value of $h_1'\in[0,h]$ such that $f(R[0,h_1'])\ge d_1$. Some $h_1'$ exists because $h_1'=h$ gives $f(R[0,h_1'])=f(R)\ge d_1$, and the minimum exists since the interval of $h_1'$ that work is closed. 
Since $f(\cdot)$ is supported on finitely many points, there exists a $\delta>0$ sufficiently small that, for $h_1=h_1'+\delta$, we have $f(R[0,h_1]) = f(R[0,h_1'])$. In particular, $R[0,h_1]$ has no support of $f(\cdot)$ on its upper edge.

\begin{enumerate}
  \item \textbf{Case 1: $h_1 \le 10\ell$.}
    In this case, $f(R[10\ell,h])\le f(R[h_1,h])=f(R)-f(R_0) \le f(R)-d_1$.
    Apply the induction hypothesis to $R[10\ell,h]$ to get a division $R_1,\dots,R_m$.
    Let $R_0=R[0,10\ell]$.
    We claim $R_0,R_1,\dots,R_m$ is our desired division of $R$. Rectangles $R_1,\dots,R_m$ satisfy properties 1 and 2 by induction, and rectangle $R_0$ satisfies properties 1 and 2 as the height is $10\ell$.
    The number of rectangles is
    \begin{align}
  1+m \le 1+\frac{2f(R[10\ell,h])}{d_1} \le 1 + \frac{2(f(R) - d_1)}{d_1} < \frac{2f(R)}{d_1}.
    \end{align}
      Thus, this is a valid division.

    \item \textbf{Case 2: $h_1 > 10\ell$.}
      In this case, $f(R[h_1,h])=f(R)-f(R[0,h_1]) \le f(R)-d_1$.
      Apply the induction hypothesis to $R[h_1,h]$ to get a division $R_1,\dots,R_{m}$.
      Let $R_{0,0}=R[0,h_1-5\ell]$ and $R_{0,1}=R[h_1-5\ell, h_1]$.
      We claim $R_{0,0},R_{0,1},R_1,\dots,R_m$ is our desired division of $R$.
      Rectangles $R_1,\dots,R_m$ satisfy properties 1 and 2 by induction.
      Rectangles $R_{0,0}$ and $R_{0,1}$ satisfy properties 1 and 2 by definition: for property 1, both have height at least $5\ell$, and for property 2, we have $f(R_{0,0}) < d_1$ by minimality of $h_1'$ (we use $h_1-5\ell < h_1'$) and $R_{0,1}$ has height $5\ell$.
      The number of rectangles is
      \begin{align}
    2+m \le 2+\frac{2f(R[h_1,h])}{d_1} \le 2 + \frac{2(f(R) - d_1)}{d_1} = \frac{2f(R)}{d_1}.
      \end{align}
      Thus, this is a valid division.
\end{enumerate}
This completes the casework, completing the induction, completing the proof.
\end{proof}

\subsection{Proof of Theorem~\ref{thm:main-3}}
  \begin{figure}
  \begin{center}
      \begin{tikzpicture}[scale=0.45]
\draw[white,pattern=north west lines,pattern color=blue!40] (0,0) rectangle (15,15);
    \foreach \i in {0,5,...,15} {
      \draw [pattern=crosshatch,pattern color=red!100] (\i-0.2,0-0.2) rectangle (\i+0.2,15+0.2);
      \draw [pattern=crosshatch,pattern color=red!100] (0-0.2,\i-0.2) rectangle (15+0.2,\i+0.2);
    }
    \foreach \i in {0,5,...,15} {
      \foreach \j in {0,5,...,15} {
        \draw[fill=yellow!90] (\i-0.4,\j-0.4) rectangle (\i+0.4,\j+0.4);
      }
    }
    \draw [pattern=crosshatch,pattern color=red!100] (0-0.2,7-0.2) rectangle (5+0.2,7+0.2);
    \draw[fill=yellow!90] (0-0.4,7-0.4) rectangle (0+0.4,7+0.4);
    \draw[fill=yellow!90] (5-0.4,7-0.4) rectangle (5+0.4,7+0.4);

    \draw [pattern=crosshatch,pattern color=red!100] (10-0.2,13-0.2) rectangle (15+0.2,13+0.2);
    \draw[fill=yellow!90] (10-0.4,13-0.4) rectangle (10+0.4,13+0.4);
    \draw[fill=yellow!90] (15-0.4,13-0.4) rectangle (15+0.4,13+0.4);
    \node at (7.5,19) {\textbf{Case 1: $k\ge d$}};
    \node (a) at (1,17) {$\mathcal{A}$};
    \draw[->] (a) to[out=300,in=90] (2.5,12.5);

    \node (a) at (10,17) {Bad Rect};
    \draw[->] (a) to[out=330,in=90] (12,14);
    \draw[->] (a) to[out=330,in=90] (13,11.5);
    \draw[<->] (-0.7,10) -- (-0.7,15) node [midway, left] {$w$};
    \draw[<->] (-0.7,5-0.4) -- (-0.7,5+0.4) node [midway, left] {$\Theta(\ell)$};
  \end{tikzpicture} 
  \quad
      \begin{tikzpicture}[scale=0.45]
\draw[white,pattern=north west lines,pattern color=blue!40] (0,0) rectangle (15,15);
    \foreach \i in {0,5,...,15} {
      \draw [pattern=crosshatch,pattern color=red!30] (\i-0.4,-0.4) rectangle (\i+0.4,15.4);
      \draw [pattern=crosshatch,pattern color=red!30] (0-0.4,\i-0.4) rectangle (15+0.4,\i+0.4);
      \draw [pattern=crosshatch,pattern color=red!100] (\i-0.2,0-0.2) rectangle (\i+0.2,15+0.2);
      \draw [pattern=crosshatch,pattern color=red!100] (0-0.2,\i-0.2) rectangle (15+0.2,\i+0.2);
    }
    \foreach \i in {0,5,...,15} {
      \foreach \j in {0,5,...,15} {
        \draw[fill=yellow!90] (\i-0.4,\j-0.4) rectangle (\i+0.4,\j+0.4);
      }
    }
    \node[] at (7.5,19) {\textbf{Case 2: $d\ge k$}};
    \node (b) at (1,17) {$\mathcal{B}$};
    \node (b') at (3,17) {$\mathcal{B}'$};
    \node (c) at (8,17) {$\mathcal{C}$};

    \draw[->] (b) to[out=300,in=90] (2,15);
    \draw[->] (b') to[out=270,in=90] (3.5,15.4);
    \draw[->] (c) to[out=270,in=90] (10,15);
  \end{tikzpicture} 
  \end{center}
  \caption{Our division of the plane into $\mathcal{A}$ (lined blue), $\mathcal{B}$ (crosshatch dark red crosshatch), and $\mathcal{C}$ (solid yellow), along with the region $\mathcal{B}'$ (crosshatch dark and light red).
  These regions inform our qubit division $Q=A\sqcup B\sqcup C$.
  We use this division in different ways for the cases $k\ge d$ and $d\ge k$.
  When $k\ge d$ (Left), we ignore $\mathcal{B}'$, and also subdivide bad squares into bad rectangles.
  When $d\ge k$ (Right), there are no bad squares, but we use the region $\mathcal{B}'$.
  }
  \label{fig:abc}
  \end{figure}
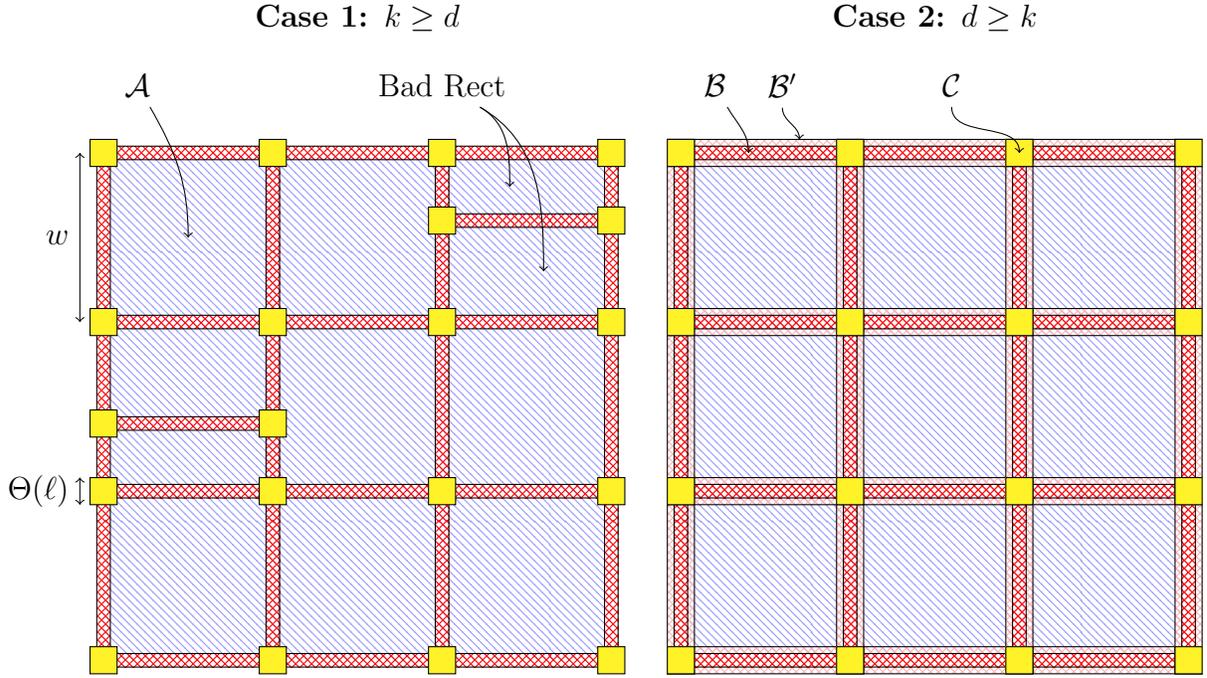
In this section, we prove Theorem~\ref{thm:main-3}.
\begin{proof}[Proof of Theorem~\ref{thm:main-3}]
    We give a proof by contradiction. We assume for contradiction our stabilizer code has few long interactions, and then obtain a contradiction by showing the dimension is less than $k$. Specifically, we find a partition of the qubits $A\sqcup B\sqcup C$ such that $A$ and $B$ are correctable and $|C|<k$, and then use Lemma~\ref{lem:bptabc}.
    
    Choose $c_0=\frac{1}{100}$ and $c_1=c_0^{-4}$.
    Assume we have an $[[n,k,d]$ stabilizer code with a 2D-embedding $Q\subset\mathbb{R}^2$ satisfying $kd^2 \geq c_1 \cdot n$, and let $\ell=c_0\big( \frac{kd^2}{n} \big)^{1/4}$.
    By assumption, $1\le \ell\le \frac{\sqrt{d}}{100}$.
    We claim that this embedding has at least $c_0\cdot \max(k,d)$ interactions of length $\ge\ell$.
    If $d\ge \sqrt{kn}$, the result follows from Theorem~\ref{thm:main-2} (with a slightly worse $c_0$), so we may assume that $d\le \sqrt{kn}$.
    
    With hindsight, let $w=d/{100\ell}$.
    Suppose for contradiction that we have a 2D-local $[[n,k,d]]$ stabilizer code with at most $\max(k,d)/100$ interactions of length $\ge \ell$.

  Let $Q\subset\mathbb{R}^2$ be the 2D-embedding of qubits.
  Call an interaction \emph{good} if the length of the interaction is less than $\ell$, and \emph{bad} otherwise.
  Call a qubit $v\in Q$ \emph{bad} if it participates in a bad interaction, and \emph{good} otherwise.
  With these definitions, the function $f_{\ge\ell}(v)$ counts the number of bad interactions that the qubit $v\in Q$ participates in.
  Clearly, $f_{\ge\ell}$ is finitely supported, and, by assumption, $\sum_{v\in Q} f_{\ge\ell}(v) \le \max(k,d)/50$, so there are at most $\max(k,d)/50$ bad qubits.

  We now construct a division of the plane $\mathcal{A}\sqcup\mathcal{B}\sqcup\mathcal{C}$ that informs our partition of the qubits $Q=A\sqcup B\sqcup C$ (see Figure~\ref{fig:abc}).
  First apply Lemma~\ref{lem:square-2} with $X=Q$ and $Y$ is the multiset of qubits where qubit $v$ appears with multiplicity $f_{\ge\ell}(v)$.
  This gives a tiling of $\mathbb{R}^2$ with $w\times w$ squares $\{S_{i,j}\}_{i,j\in\mathbb{Z}}$ where at most $\frac{32\ell^2}{w^2}\cdot n$ qubits of $Q$ are within distance $2\ell$ of some vertex of some square, and at most $\frac{16\ell}{w}\cdot \max(k,d)/50$ bad interactions involve a qubit within distance $2\ell$ of some edge of some square.
  Call a square $S_{i,j}$ \emph{good} if $f_{\ge\ell}(S_{i,j})< d/10$ and \emph{bad} otherwise.
For each bad square $S_{i,j}$, take a decomposition $R_{i,j,1},\dots,R_{i,j,m_{i,j}}$ given by Lemma~\ref{lem:square-3}. 
Call all the rectangles $R_{i,j,r}$ \emph{bad}.
In this way, the plane is partitioned into good squares and bad rectangles.
The total number of bad rectangles is at most
\begin{align}
  \sum_{i,j:\text{$S_{i,j}$ bad}}^{} 2\cdot\frac{f_{\ge\ell}(S_{i,j})}{d/10} 
  \le \sum_{i,j}^{} 2\cdot \frac{f_{\ge\ell}(S_{i,j})}{d/10} 
  < \frac{\max(k,d)}{d}.
  \label{eq:num-r}
\end{align}
where we used that $\sum_{i,j}^{} f_{\ge\ell}(S_{i,j})\le \max(k,d)/50$.
Define the partition $\mathcal{A},\mathcal{B},\mathcal{C}$ of the plane as follows (see Figure~\ref{fig:abc}):
\begin{itemize}
  \item $\mathcal{C}$ is the set of all  points within $\ell_\infty$-distance $2\ell$ of some vertex of some $R_{i,j,r}$ or $S_{i,j}$.

  \item $\mathcal{B}$ is the set of all points both (1) not in $\mathcal{C}$ and (2) within distance $\ell$ of some edge of a good square $S_{i,j}$ or bad rectangle $R_{i,j,r}$.
  \item $\mathcal{B}'$ is the set of all points both (1) not in $\mathcal{C}$ and (2) within distance $2\ell$ of some edge of a good square $S_{i,j}$.

  \item $\mathcal{A}$ is the set of points not in $\mathcal{B}$ or $\mathcal{C}$.
\end{itemize}
By slightly perturbing the tiling, we may assume without loss of generality that no points lie on the boundary of any region of $\mathcal{A},\mathcal{B},$ or $\mathcal{C}$.

We now derive a contradiction by finding sets $A,B,C$ of qubits such that $A$ and $B$ are correctable and $|C|<k$.
We need two cases, depending on which of $k$ or $d$ is larger.
The two cases come from two different techniques for handling the bad interactions in our partition $A\sqcup B\sqcup C$. When the number of bad interactions is less than $k/100$, we simply put the bad qubits in $C$. 
When the number of bad interaction is less than $d/100$, the set of bad qubits is correctable. Since we chose the initial grid tiling carefully, almost all bad qubits don't interact with $B$, so by the Union Lemma we can add almost all bad qubits to $B$ while keeping $B$ correctable.

\paragraph{Case 1: $k\ge d$.}

Define the partition $A,B,C$ of the qubits $Q$ as follows:
\begin{itemize}
  \item $C$ is the set of all of the following qubits:
    \begin{itemize}
      \item qubits in region $\mathcal{C}$
      \item all bad qubits.
    \end{itemize}

  \item $B$ is the set of all good qubits in region $\mathcal{B}$.

  \item $A$ is the set of all good qubits in region $\mathcal{A}$.
\end{itemize}
It is easy to check that this partitions the qubits $Q$.
We now check that $A$ and $B$ are correctable and $|C|<k$, giving the desired contradiction.
\begin{itemize}
  \item \textbf{$A$ is correctable.} 
    The entire plane is decomposed into good squares and bad rectangles.
    Arbitrarily enumerate these regions (the good squares and bad rectangles) $\mathcal{A}_1',\mathcal{A}_2',\dots$ (we use the notation $\mathcal{A}_i'$ rather than $\mathcal{A}_i$ to clarify that $\mathcal{A}_i'$ partition the plane, not $\mathcal{A}$).
    Let $A_i'$ denote the set of qubits in $\mathcal{A}_i'$.
    If $\mathcal{A}_i'$ is a good square, then $A_i'$ is correctable by Lemma~\ref{lem:square-1}.
    If $\mathcal{A}_i'$ is a bad rectangle, then either (i) $f_{\ge\ell}(\mathcal{A}_i')\le d/10$ or (ii) $\mathcal{A}_i$ has area at most $w\cdot 10\ell=d/10$. 
    In the first case, $A_i'$ is correctable by Lemma~\ref{lem:square-1}, and in the latter, $|A_i'|<6d/10$ by Lemma~\ref{lem:square-0}, so $A_i'$ is correctable by the Distance Property (Lemma~\ref{lem:correctable}).
    Hence, $A_i'$ is correctable for all $i$.
    By Subset Closure (Lemma~\ref{lem:correctable}) $A_i\defeq A_i'\cap A$ is correctable for all $i$.
    Further, by definition of $\mathcal{A}$, the qubits in $A_i$ are at distance at least $2\ell$ from any other $A_j$.
    Since each $A_i$ consists only of good qubits, the qubit sets $A_1,A_2,\dots$ are pairwise decoupled, so by the Union Lemma (Lemma~\ref{lem:correctable}), $A=\cup_i A_i$ is correctable.

  \item \textbf{$B$ is correctable.} The region $\mathcal{B}$ can be decomposed in rectangles with one side length equal to $2\ell$ and other side length at most $w-2\ell$. Enumerate these rectangles arbitrarily as $\mathcal{B}_1,\mathcal{B}_2,\dots$.
    Let $B_i$ be the set of good qubits in $\mathcal{B}_i$.
    By Lemma~\ref{lem:square-0}, $B_i$ has at most $6\cdot 2\ell(w-2\ell) < d$ qubits, so $B_i$ is correctable for all $i$ by the Distance Property (Lemma~\ref{lem:correctable}).
    Further, any two regions $\mathcal{B}_i$ are at distance at least $\ell\sqrt{2}$ (see Figure~\ref{fig:B}).
    Since each $B_i$ consists only of good qubits, we have that $B_1,B_2,\dots$ are pairwise decoupled, so by the Union Lemma (Lemma~\ref{lem:correctable}), $B=\cup_i B_i$ is correctable.

\begin{figure}
    \centering
    \begin{tikzpicture}[scale=1.5]
      \draw [fill=red!80,pattern=crosshatch, pattern color=red,draw=none] (0-0.5,2) rectangle (0+0.5,5);
      \draw [fill=red!80,,pattern=crosshatch, pattern color=red,draw=none] (4-0.5,2) rectangle (4+0.5,5);
      \draw [fill=red!80,pattern=crosshatch, pattern color=red,draw=none] (0,5-0.5) rectangle (4,5+0.5); 
    \foreach \i in {0,4} {
      \foreach \j in {5} {
        \draw[fill=yellow!90,draw=none] (\i-1,\j-1) rectangle (\i+1,\j+1);
      }
    }
    \draw[<->,dotted] (0.5,3) -- (3.5,3) node [midway, above] {\Large $\ge3\ell$};
    \draw[<->,dotted] (0,5) -- (0,4) node [midway, left] {$2\ell$};
        \draw[<->,dotted] (0,5) -- (1,5) node [midway, above] {$2\ell$};
    \draw[<->,dotted] (0,4) -- (0.5,4) node [midway, below] {$\ell$};
    \draw[<->,dotted] (0,4) -- (-0.5,4) node [midway, below] {$\ell$};
    \draw[<->,dotted] (0.5,4) -- (1,4.5);
    \node at (1.1,4.1) {\Large $\ell\sqrt{2}$};
    \node at (0,3) {\Large $\mathcal{B}_i$};
    \node at (4,3) {\Large $\mathcal{B}_{i'}$};
    \node at (2,5) {\Large $\mathcal{B}_{i''}$};
    \node at (4,5) {\Large $\mathcal{C}$};
    \end{tikzpicture}
\caption{The region $\mathcal{B}$ can be partitioned via edge-centered rectangles (red regions). By definition, any two perpendicular edge-centered rectangles are at distance at least $\ell\sqrt{2}$, and any two edge-centered parallel rectangles are at distance at least $3\ell$.}
    \label{fig:B}
\end{figure}
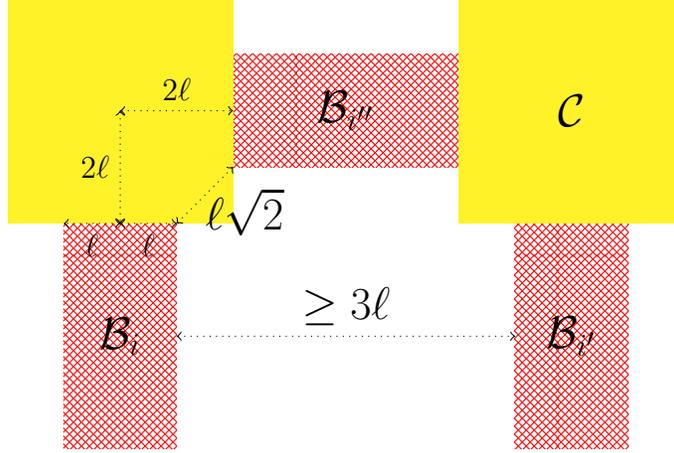

\item \textbf{$|C| < k$.} The number of bad qubits is at most $k/50$ by assumption.
  The number of qubits within $\ell_\infty$-distance $2\ell$ of the vertex of a square $S_{i,j}$ is at most $\frac{16\ell^2}{w^2}n$ by our choice of $S_{i,j}$.
The number of bad rectangles is less than $\frac{k}{d}$ by \eqref{eq:num-r}. 
The number of qubits within $\ell_\infty$-distance $2\ell$ of some particular vertex of some $R_{i,j,r}$ is at most $6\cdot 16\ell^2$ by Lemma~\ref{lem:square-0}.
Thus, the number of points within $\ell_\infty$ distance $2\ell$ of some bad rectangle's vertex is less than $\frac{400\ell^2k}{d}$.
Thus, 
\begin{align}
  |C|
  \le \frac{k}{50} + \frac{400\ell^2k}{d} + \frac{16\ell^2n}{w^2} 
  &\le \frac{k}{50} + \frac{k}{25} +\frac{k}{100}  < k.
\end{align}
In second inequality, the middle term plugs in $\ell$ and uses $k\le n$, and the last term plugs in $w$ and $\ell$ in terms of $n,k,d$ and simplifies. 
\end{itemize}

\paragraph{Case 2: $d\ge k$.}

By \eqref{eq:num-r}, there are no bad rectangles, only good squares (the number of bad rectangles is less than 1).
Define the partition $A,B,C$ of the qubits $Q$ as follows:
\begin{itemize}
  \item $C$ is the set of all of the following qubits:
    \begin{itemize}
      \item qubits in region $\mathcal{C}$
      \item qubits participating in a bad interaction with a qubit in $\mathcal{B}'$ (including bad qubits in $\mathcal{B}'$).
    \end{itemize}

  \item $B$ is the set of all of the following qubits:
    \begin{itemize}
      \item good qubits in region $\mathcal{B}$.
      \item bad qubits not already in $C$.
    \end{itemize}

  \item $A$ is the set of all good qubits in region $\mathcal{A}$.
\end{itemize}

It is easy to check that this partitions the qubits on $Q$.
We now check this partition has the desired properties: $A$ and $B$ are correctable and $|C|<k$, giving the desired contradiction.
\begin{itemize}
  \item \textbf{$A$ is correctable.} 
    Since every square $S_{i,j}$ is good, the set of qubits in $S_{i,j}$ is correctable by Lemma~\ref{lem:square-1}.
    For all $i,j$, let $A_{i,j}$ be the set of qubits of $A$ in region $S_{i,j}$.
    By Subset Closure (Lemma~\ref{lem:correctable}), $A_{i,j}$ is also correctable.
    Further, since $A$ consists only of good qubits, and any two qubits $A_{i,j}$ and $A_{i',j'}$ are at distance at least $2\ell$, any two qubit sets $A_{i,j}$ and $A_{i',j'}$ are decoupled.
    Since $A=\cup_{i,j}A_{i,j}$, we have that $A$ is correctable by the Union Lemma (Lemma~\ref{lem:correctable}).
  
  \item \textbf{$B$ is correctable.}
    The region $\mathcal{B}$ consists of $2\ell\times (w-2\ell)$ rectangles. Enumerate these regions $\mathcal{B}_1,\mathcal{B}_2,\dots,$ arbitrarily, and let $B_i$ be the set of qubits in set $B$ and in region $\mathcal{B}_i$.
    By Lemma~\ref{lem:square-0}, each $\mathcal{B}_i$ has at most $12\ell(w-2\ell)<d$ qubits, so $B_i$ is correctable by the Distance property. 
    Further, $B_1,B_2,\dots,$ are pairwise decoupled because $B_1,B_2,\dots$ do not contain any bad qubits by construction, and any two of $B_i$ and $B_j$ are at distance at least $\ell\sqrt{2}$ (see Figure~\ref{fig:B}).
    Let $B_0$ be the set of bad qubits not already in $C$.
    By assumption, the number of bad interactions is at most $d/100$, so $|B_0|\le d/50$ and $B_0$ is correctable by the Distance Property (Lemma~\ref{lem:correctable}).

    By construction, the qubits in $B_0$ do not have any bad interactions with the qubits in $\mathcal{B}$, and thus with $B_1,B_2, \dots $. Further, since the qubits in $B_0$ are not in $\mathcal{B}'$, the qubits in $B_0$ have distance at least $\ell$ from region $\mathcal{B}$, and thus no good interactions with $B_1,B_2,\dots$ either. Hence, $B_0$ is decoupled from $B_1,B_2,\dots$.
    Thus, by the Union Lemma, $B=B_0\cup B_1\cup B_2\cup\cdots$ is correctable.

  \item \textbf{Size of $C$:} The number of qubits in $\mathcal{C}$ is at most $\frac{16\ell^2}{w^2}n$ by our choice of $S_{i,j}$. By the definition of our initial tiling, the total number of bad interactions with $\mathcal{B}'$ is at most $\frac{16\ell}{w}\cdot \frac{d}{50} < \frac{\ell}{2w}\cdot d$, so there are at most $\frac{\ell}{w}\cdot d$ qubits participating in a bad interaction with a qubit in $\mathcal{B}'$. 
    Hence,
    \begin{align}
    |C|  \le \frac{16\ell^2}{w^2}\cdot n + \frac{\ell}{w}\cdot d 
    \le \frac{k}{100} + \frac{k}{100} < k 
      \label{}
    \end{align}
    In the first middle inequality, we used the definitions of $w$ and $\ell$ in terms of $n,k,d$, and then used $d\le \sqrt{kn}$ for the second term.
\end{itemize}

This completes the casework, completing the proof.
    \end{proof}

\section{Optimality}
\label{sec:construction}

\subsection{Theorem~\ref{thm:good}, Interaction Count is Optimal}
For this section we recall the following basic \emph{padding} operation on codes (see, e.g., \cite{gottesman1997stabilizer}). Given an $[[n,k,d]]$ stabilizer code, for any integer $r\ge 1$, one can construct an $[[n+r,k,d]]$ stabilizer code with the same number of interactions as the original code. 
For $r=1$, the new stabilizer contains the generator $\ssI^{\otimes n} \otimes \ssX$, along with additional generators $M\otimes \ssI$ for every generator $M$ in the original code. 
It is easy to check this new code has the desired parameters, and the new ancillary qubit does not add any new interactions.
This easily generalizes to $r>1$.

\begin{proof}[Proof of Theorem~\ref{thm:good}] 
  By \cite{panteleev2021asymptotically}, for all $n,k,d$, there is an $[[n_1, \Omega(k),\Omega(d)]]$ qLDPC code for some $n_1 = \Theta(\max(k,d)) \le n$.
  We briefly justify this statement, as \cite{panteleev2021asymptotically} does not explicitly claim good codes for \emph{all} blocklengths $n$.
  By standard facts about the distribution of primes, it follows by inspection of \cite[Theorem 2]{panteleev2021asymptotically} that, for some absolute constant $C$, for any sufficiently large $n_2\ge C$, there exists an qLDPC code of length $n_1\in[0.99n_2,n_2]$ dimension $\Omega(n_1)$ and distance $\Omega(n_1)$.
  \footnote{
      For completeness, we work it out more explicitly. The code in \cite[Theorem 2]{panteleev2021asymptotically} is determined by a rate $R\in(0,1)$. This rate $R$ determines constants $r,r'$ and a fixed prime $p$ that is 1 mod 4 (in their notation, $p=w-1$). Their code length is $n=\frac{1}{2}n_1(t)((p+1)^2+4rr')$, for some $n_1(t)=t(t^2-1)$, where $t$ is a sufficiently large prime that is 1 mod 4 satisfying $p^{(t-1)/2}\equiv 1 \mod t$. In particular, we may take $t$ to be any sufficiently large prime that is 1 mod $4p$: such a $t$ is clearly 1 mod 4 and 1 mod $p$, which implies $t$ is a quadratic residue mod $p$, which, by quadratic reciprocity, implies that $p$ is a quadratic residue mod $t$, which implies that $p^{(t-1)/2}\equiv 1\mod t$. By the Prime Number Theorem for Arithmetic Progressions (see, e.g., \cite{mccurley1984explicit}), there is a prime $t\equiv 1\mod 4p$ in every interval $[0.999n_2, n_2]$ for all sufficiently large $n_2$, so there is a valid code length in every interval $[0.99n_2,n_2]$ for all sufficiently large $n_2$.
      We note that ``sufficiently large" depends only on $R$, which we can take here to simply be $1/2$.
  }

   Choose $n_2=\max(k,d)$. We may assume $n_2\ge C$ or else any code satisfies the theorem (if $n_2\le C=O(1)$, then $1$ is $\Omega(k)$ and $\Omega(d)$). 
   There is a length $n_1\in [0.99 n_2, n_2]$ qLDPC code with dimension $\Omega(k)$ and distance $\Omega(d)$.
   Recall that a qLDPC code's stabilizer has a set of generators all of which are supported on $O(1)$ qubits.
   Thus, each stabilizer contributes $O(1)$ interactions, so there are $O(\max(k,d))$ interactions total. 
    We now pad the code with $n-n_1$ qubits, to obtain a code with length at most $n$, dimension at least $\Omega(k)$, distance at least $\Omega(d)$, and the same number of interactions, $O(\max(k,d))$.
\end{proof}

\subsection{Theorem~\ref{thm:good}, Interaction Length is Optimal}
To show that our bounds on the code parameters are optimal, we show the existence of quantum codes that match these bounds in strong ways. Our construction relies on a variation of quantum code concatenation \cite{gottesman1997stabilizer}. Informally, if we have a $[[n_1,k_1,d_1]]$ code $\mathcal{S}_1$ (the inner code)  and a $[[n_2,k_2,d_2]]$ code $\mathcal{S}_2$ (the outer code), we perform the following procedure: 
\begin{enumerate}
    \item Take $n_2$ blocks of the inner code $\mathcal{S}_1$
    \item For each block, encode $k_1$ logical qubits using the $n_1$ physical qubits in each block
    \item Take $k_1$ copies of the outer code $\mathcal{S}_2$.
    \item Replace the $j$-th physical qubit in the $i$-th copy of $\mathcal{S}_2$ by the $i$-th logical qubit encoded in the $j$-th $\mathcal{S}_1$ code block. 
\end{enumerate}
The concatenated code $\mathcal{S}$ contains $k_1$ copies of $\mathcal{S}_2$, so it encodes $k_1k_2$ logical qubits. The distance is at least $d_1d_2$, since there must be at least $d_1$ errors in $d_2$ blocks of $\mathcal{S}_1$ to cause an undetectable error for the whole code.
When the inner and outer codes are stabilizer codes, the resulting code is also a stabilizer code, and its stabilizer is constructed in the following way:

\begin{lemma}[Code Concatenation, see e.g., Section 3.5 of \cite{gottesman1997stabilizer}]
\label{thm:copy-concatenation}
    Given stabilizer codes $[[n_1,k_1,d_1]]$ and $[[n_2,k_2,d_2]]$ with stabilizers $\mathcal{S}_1, \mathcal{S}_2$, there exists a $[[n_1n_2, k_1k_2, d \geq d_1d_2]]$ code $C$ with stabilizer $\mathcal{S}$, where $\mathcal{S}$ is generated by operators that are either
    \begin{enumerate}
        \item the tensor product of a stabilizer generator $M$ from one of $n_2$ code blocks of $\mathcal{S}_1$ with the identity operator on all other such code blocks.         \item an operator $\overline{M}$ formed by taking a stabilizer generator $M$ from the $j$-th copy of $\mathcal{S}_2$, and replacing the Pauli operator $P_i$ (acting on qubit $i$) from its tensor product decomposition with the corresponding logical Pauli operator $\overline{P}_{ij}$ for the $j$-th logical qubit encoded through the $i$-th $\mathcal{S}_1$ code block.
    \end{enumerate}
 \end{lemma}

With Lemma~\ref{thm:copy-concatenation} and Theorem~\ref{thm:good} in hand, we can now give a construction of a code showing the optimality of Theorem \ref{thm:main}. The central idea is that although good qLDPC codes ``optimally" use the number of physical qubits to produce the largest possible rate and distance, they do not provide any way of controlling the interaction length. The surface code, due to its locally interacting generators, provides a way to control the interaction length. By concatenating an inner good code with an outer surface code, we replace the surface code qubits with good code blocks, thus allowing us to utilize the efficiency of good codes while confining their interactions to a restricted space. We remark that our concatenated code is \emph{not} a qLDPC code, since the logical operators in $\mathcal{S}_2$ have high weight. Nonetheless, we present the code to show that it saturates our bound on the interaction length.

We additionally note that concatenation alone does not saturate the bound everywhere. Specifically, this concatenation saturates the bounds in Theorem~\ref{thm:main} for all $d\sim\sqrt{kn}$ (and consequently, all $d\ge \sqrt{kn}$).
When $d\ll \sqrt{kn}$, we take a optimal concatenated construction of length $n' \sim d^2/\ell^2$ with dimension $k'\sim\ell^2$, distance $d$ (so that $d\sim\sqrt{k'n'}$), and max interaction length $\ell$, and then copy that construction $n/n'$ times, so that the overall dimension is $k'\cdot n/n' = k$, overall length is $n$, overall distance is still $d$, and overall max interaction length is still $\ell$.
This allows us to saturate the bound everywhere.

\begin{theorem}[Construction saturating the bound]
 There exists an absolute constants $c_0,c_1\ge 1$ such that the following holds.
   For all $n,k,d$ with $k,d\le n$ and $kd^2\ge c_1\cdot n$, there exists a $[[n,\ge k/c_0,\ge d/c_0]]$ quantum stabilizer code with \emph{no} interactions of length at least $\ell=c_0\cdot\max\big(\frac{d}{\sqrt{n}}, \big( \frac{kd^2}{n} \big)^{1/4} \big)$.
   
\end{theorem}

\begin{proof}
We ignore rounding errors from non-integer parameters; as long as all numbers we work with are at least 1, these incur at most a constant factor over the final constants and does not change the final statement.
We can indeed check that all the parameters below that ``should'' be integers are at least 1 --- in particular, $n', n/n', \varepsilon n_1, n_2,\ell/10 \ge 1$ --- so that cost of rounding overall costs at most a constant factor.

It suffices to consider the case $d\le \sqrt{kn}$. If $d > \sqrt{kn}$, then take $k'\ge k$ such that $d=\sqrt{k'n}$.
The $d\le \sqrt{k'n}$ case implies the existence of a code with no interactions of length at least $c_0\cdot \max\big(\frac{d}{\sqrt{n}}, \big( \frac{k'd^2}{n} \big)^{1/4} \big) = c_0\cdot \frac{d}{\sqrt{n}}$, as desired.

By Theorem \ref{thm:good} \cite{panteleev2021asymptotically,leverrier2022quantum}, there exists a good qLDPC code $\mathcal{S}_1$ with parameters $[[n_1, k_1 \geq \varepsilon n_1, d_1 \geq \varepsilon n_1]]$, for some $\varepsilon \in (0,1)$. With hindsight, let $c_0=100/\varepsilon$.
Now define $\ell=c_0\cdot\max\big(\frac{d}{\sqrt{n}}, \big( \frac{kd^2}{n} \big)^{1/4} \big)$, and choose $n' = d^2/\ell^2$ so that $n/n' = n\ell^2/d^2$.
Choose $n_1=\ell^2/100$ and $n_2 = n'/n_1$.

Let $\mathcal{S}_2$ be a surface code with parameters $[[n_2, k_2=1, d_2=\sqrt{n_2}]]$ \cite{horsman2012surface}. 
Applying Lemma~\ref{thm:copy-concatenation}, we concatenate the inner code $\mathcal{S}_1$ with the outer code $\mathcal{S}_2$ to obtain a code $\mathcal{S}'$ with parameters $[[n', k' \geq \varepsilon n_1, d' \geq \varepsilon n_1\cdot \sqrt{n_2}]]$.  
Take $n/n'$ copies of $\mathcal{S}'$ to get a new code $\mathcal{S}$ using $n$ qubits with dimension 
\begin{equation}
    k' \cdot \frac{n}{n'} \geq \frac{\varepsilon n\ell^4}{100d^2} \geq \frac{\varepsilon k}{100}
\end{equation}
and distance
\begin{equation}
    d' \geq \frac{d}{10}.
\end{equation}

It remains to show that the concatenated code $\mathcal{S}'$ has a 2D-embedding with no interactions of length at least $\ell$. 
The 2D-embedding of the final code $\mathcal{S}$ follows by arranging the $n/n'$ embedding copies of $\mathcal{S'}$  disjointly in the plane. 
First, we give the 2D-embedding of $\mathcal{S}'$. We start with a $\sqrt{n_2} \times \sqrt{n_2}$ lattice, where each edge on the lattice has length $\frac{4}{10}\ell$. On each vertex, we place a $(\frac{\ell}{10})\times(\frac{\ell}{10})$ square, upon which we embed the $n_1$ qubits in each code block of $\mathcal{S}_1$. Note that the maximum distance between two qubits contained in the same plaquette is then at most 
\begin{equation}
    \label{eqn:max-interaction-length}
   \left(\frac{1}{2}\cdot \frac{1}{10}\ell + \frac{4}{10}\ell + \frac{1}{2}\cdot \frac{1}{10}\ell\right)\cdot \sqrt{2} < \ell. 
\end{equation}

This is the 2D-embedding of the concatenated code $\mathcal{S}'$, which is really just the usual 2D-embedding of a surface code with boundary, with the qubits replaced by the logical qubits of the qLDPC code blocks. Note that the embedding essentially layers each of the $k_1$ copies of the surface code $\mathcal{S}_1$ together. See figure \ref{fig:construction} for a sketch with important lengths labeled. 

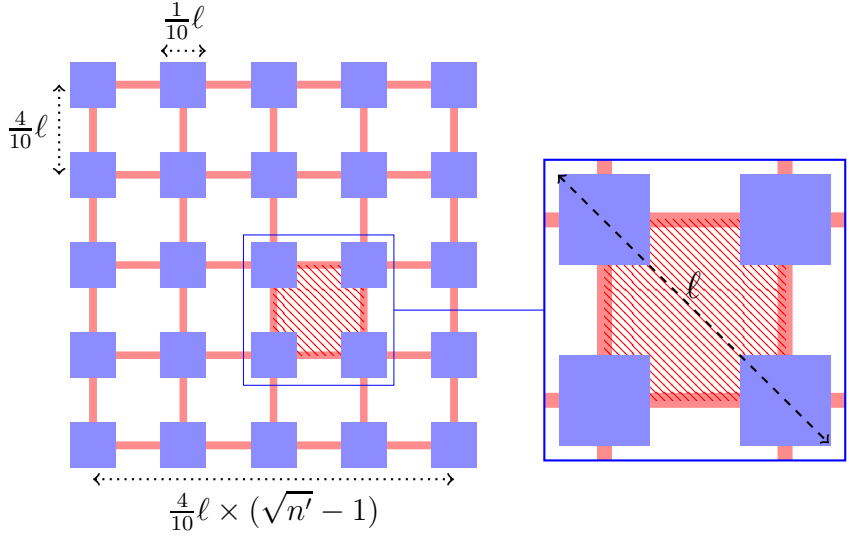
\begin{figure}
    \centering
    \begin{tikzpicture}[scale=0.3,spy using outlines={rectangle, magnification=2, size=4cm, connect spies}]   

    \draw[line width=0.1cm, red!45] (0,0) rectangle (16,16);
    \foreach \i in {0,4,...,16} {
      \draw [line width=0.1cm,red!45] (\i,0) -- (\i,16);
      \draw [line width=0.1cm,red!45] (0,\i) -- (16,\i);
    }

    \filldraw [draw=none,pattern color=red,pattern=north west lines] (8,4) rectangle (12,8);

    \foreach \i in {0,4,...,16} {
        \foreach \j in {0,4,...,16} {
            \draw[draw=none,fill=blue!45] (\i-1,\j-1) rectangle (\i+1,\j+1);
        }
    }
    \draw[<->,dotted,thick] (-1.5,12) -- (-1.5,16) node [midway, left] {$\frac{4}{10}\ell$};
    \draw[<->,dotted,thick] (3,17.5) -- (5,17.5) node [midway, above] {$\frac{1}{10}\ell$};
    \draw[<->,dotted,thick] (0,-1.5) -- (16,-1.5) node [midway, below] {$\frac{4}{10}\ell \times (\sqrt{n'}-1)$};

    \coordinate (zoom) at (10,6);
    \coordinate (spy) at (20,6);
    \spy [width=4cm, height=4cm,color=blue] on (zoom) in node [right] at (spy);
    
    \begin{pgfonlayer}{fg}
    \draw[<->, dashed, thick] (20.6,12) -- (32.6,0.1) node [midway, above]{\large $\ell$};
    \end{pgfonlayer}
    
  \end{tikzpicture}
  \caption{A sketch of the 2D embedding of the concatenated code $\mathcal{S}'$. The red lines are the lattice for the rotated surface code $\mathcal{S}_2$, while the blue squares are good qLDPC code blocks $\mathcal{S}_1$. Each block of $\mathcal{S}_1$ encodes $\Theta(\ell^2)$ copies of the corresponding physical qubit of $\mathcal{S}_2$, and the farthest distance between any two qubits participating in a stabilizer measurement (shaded red square) is at most $\ell$. }

  \label{fig:construction}
\end{figure}

Now we show that the 2D-embedding of $\mathcal{S}$ given contains no interactions of length at least $\ell$. First, consider stabilizer generators of type (1) in Lemma~\ref{thm:copy-concatenation}. Since the only qubits interacting belong to the same code block of $\mathcal{S}_1$, the interaction length is bounded above by the length of the diagonal across the block, which is $\frac{1}{10}\ell \cdot \sqrt{2} < \ell$.
So there are no long interactions from generators of type (1). Now for stabilizer generators of type (2) in Theorem \ref{thm:copy-concatenation}, note that all stabilizer generators in $\mathcal{S}_2$ are supported only on qubits belonging to the same plaquette. The largest separation between two code blocks was specifically shown in \eqref{eqn:max-interaction-length} to be at most $\ell$, so there are also no long interactions from generators of type (2). These are the only types of generators in $\mathcal{S}$, so we conclude that the 2D-embedding of $\mathcal{S}$ contains no interactions of length at least $\ell$. 
\end{proof}

\section*{Acknowledgements}

We thank Eugene Tang and Ning Bao for helpful discussions and feedback on the paper. We thank Chi-Yun Hsu for helpful discussions on Theorem~\ref{thm:good}. We thank Noah Berthusen for helpful discussions about \cite{berthusen2024partial}. RL is supported by NSF grant CCF-2347371. 

\bibliographystyle{plain}
\bibliography{bib}

\begin{thebibliography}{10}

\bibitem{aharonov1997fault}
Dorit Aharonov and Michael Ben-Or.
\newblock Fault-tolerant quantum computation with constant error.
\newblock In {\em Proceedings of the twenty-ninth annual ACM symposium on Theory of computing}, pages 176--188. ACM, 1997.

\bibitem{aliferis2005quantum}
Panos Aliferis, Daniel Gottesman, and John Preskill.
\newblock Quantum accuracy threshold for concatenated distance-3 codes.
\newblock {\em Quantum Information \& Computation}, 6(2):97--165, 2006.

\bibitem{baspin2023lower}
Nou{\'e}dyn Baspin, Omar Fawzi, and Ala Shayeghi.
\newblock A lower bound on the overhead of quantum error correction in low dimensions.
\newblock {\em arXiv preprint arXiv:2302.04317}, 2023.

\bibitem{baspin2023improved}
Nou{\'e}dyn Baspin, Venkatesan Guruswami, Anirudh Krishna, and Ray Li.
\newblock Improved rate-distance trade-offs for quantum codes with restricted connectivity.
\newblock {\em arXiv preprint arXiv:2307.03283}, 2023.

\bibitem{baspin2022connectivity}
Nou{\'e}dyn Baspin and Anirudh Krishna.
\newblock Connectivity constrains quantum codes.
\newblock {\em Quantum}, 6:711, 2022.

\bibitem{baspin2022quantifying}
Nou{\'e}dyn Baspin and Anirudh Krishna.
\newblock Quantifying nonlocality: How outperforming local quantum codes is expensive.
\newblock {\em Physical Review Letters}, 129(5):050505, 2022.

\bibitem{bergeron2020silicon}
L~Bergeron, C~Chartrand, ATK Kurkjian, KJ~Morse, H~Riemann, NV~Abrosimov, P~Becker, H-J Pohl, MLW Thewalt, and S~Simmons.
\newblock Silicon-integrated telecommunications photon-spin interface.
\newblock {\em PRX Quantum}, 1(2):020301, 2020.

\bibitem{berthusen2024toward}
Noah Berthusen, Dhruv Devulapalli, Eddie Schoute, Andrew~M Childs, Michael~J Gullans, Alexey~V Gorshkov, and Daniel Gottesman.
\newblock Toward a 2d local implementation of quantum ldpc codes.
\newblock {\em arXiv preprint arXiv:2404.17676}, 2024.

\bibitem{berthusen2024partial}
Noah Berthusen and Daniel Gottesman.
\newblock Partial syndrome measurement for hypergraph product codes.
\newblock {\em Quantum}, 8:1345, 2024.

\bibitem{bluvstein2022quantum}
Dolev Bluvstein, Harry Levine, Giulia Semeghini, Tout~T Wang, Sepehr Ebadi, Marcin Kalinowski, Alexander Keesling, Nishad Maskara, Hannes Pichler, Markus Greiner, et~al.
\newblock A quantum processor based on coherent transport of entangled atom arrays.
\newblock {\em Nature}, 604(7906):451--456, 2022.

\bibitem{bombin2021interleaving}
Hector Bombin, Isaac~H Kim, Daniel Litinski, Naomi Nickerson, Mihir Pant, Fernando Pastawski, Sam Roberts, and Terry Rudolph.
\newblock Interleaving: Modular architectures for fault-tolerant photonic quantum computing.
\newblock {\em arXiv preprint arXiv:2103.08612}, 2021.

\bibitem{bombin2006topological}
Hector Bombin and Miguel~Angel Martin-Delgado.
\newblock Topological quantum distillation.
\newblock {\em Physical Review Letters}, 97(18):180501, 2006.

\bibitem{bravyi2024high}
Sergey Bravyi, Andrew~W Cross, Jay~M Gambetta, Dmitri Maslov, Patrick Rall, and Theodore~J Yoder.
\newblock High-threshold and low-overhead fault-tolerant quantum memory.
\newblock {\em Nature}, 627(8005):778--782, 2024.

\bibitem{bravyi2022future}
Sergey Bravyi, Oliver Dial, Jay~M Gambetta, Dar{\'\i}o Gil, and Zaira Nazario.
\newblock The future of quantum computing with superconducting qubits.
\newblock {\em Journal of Applied Physics}, 132(16), 2022.

\bibitem{bravyi1998quantum}
Sergey Bravyi and Alexei~Yu Kitaev.
\newblock Quantum codes on a lattice with boundary.
\newblock {\em arXiv preprint quant-ph/9811052}, 1998.

\bibitem{bravyi2010tradeoffs}
Sergey Bravyi, David Poulin, and Barbara Terhal.
\newblock Tradeoffs for reliable quantum information storage in 2{D} systems.
\newblock {\em Physical Review Letters}, 104(5):050503, 2010.

\bibitem{bravyi2009no}
Sergey Bravyi and Barbara Terhal.
\newblock A no-go theorem for a two-dimensional self-correcting quantum memory based on stabilizer codes.
\newblock {\em New Journal of Physics}, 11(4):043029, 2009.

\bibitem{breuckmann2020balanced}
Nikolas~P. Breuckmann and Jens~N. Eberhardt.
\newblock Balanced product quantum codes.
\newblock {\em IEEE Transactions on Information Theory}, pages 1--1, 2021.

\bibitem{breuckmann2021ldpc}
Nikolas~P Breuckmann and Jens~Niklas Eberhardt.
\newblock Quantum low-density parity-check codes.
\newblock {\em PRX Quantum}, 2(4):040101, 2021.

\bibitem{breuckmann2016constructions}
Nikolas~P Breuckmann and Barbara~M Terhal.
\newblock Constructions and noise threshold of hyperbolic surface codes.
\newblock {\em IEEE transactions on Information Theory}, 62(6):3731--3744, 2016.

\bibitem{calderbank1997quantum}
A~Robert Calderbank, Eric~M Rains, Peter~W Shor, and Neil~JA Sloane.
\newblock Quantum error correction and orthogonal geometry.
\newblock {\em Physical Review Letters}, 78(3):405, 1997.

\bibitem{delfosse2013tradeoffs}
Nicolas Delfosse.
\newblock Tradeoffs for reliable quantum information storage in surface codes and color codes.
\newblock In {\em Information Theory Proceedings (ISIT), 2013 IEEE International Symposium on}, pages 917--921. IEEE, 2013.

\bibitem{delfosse2021bounds}
Nicolas Delfosse, Michael~E Beverland, and Maxime~A Tremblay.
\newblock Bounds on stabilizer measurement circuits and obstructions to local implementations of quantum {LDPC} codes.
\newblock {\em arXiv preprint arXiv:2109.14599}, 2021.

\bibitem{eberhardt2024logical}
Jens~Niklas Eberhardt and Vincent Steffan.
\newblock Logical operators and fold-transversal gates of bivariate bicycle codes.
\newblock {\em arXiv preprint arXiv:2407.03973}, 2024.

\bibitem{evra2022decodable}
Shai Evra, Tali Kaufman, and Gilles Z{\'e}mor.
\newblock Decodable quantum ldpc codes beyond the n distance barrier using high-dimensional expanders.
\newblock {\em SIAM Journal on Computing}, (0):FOCS20--276, 2022.

\bibitem{flammia2017limits}
Steven~T Flammia, Jeongwan Haah, Michael~J Kastoryano, and Isaac~H Kim.
\newblock Limits on the storage of quantum information in a volume of space.
\newblock {\em Quantum}, 1:4, 2017.

\bibitem{freedman2002z2}
Michael~H Freedman, David~A Meyer, and Feng Luo.
\newblock Z2-systolic freedom and quantum codes.
\newblock {\em Mathematics of quantum computation, Chapman \& Hall/CRC}, pages 287--320, 2002.

\bibitem{fu2024error}
Xiaozhen Fu and Daniel Gottesman.
\newblock Error correction in dynamical codes.
\newblock {\em arXiv preprint arXiv:2403.04163}, 2024.

\bibitem{gottesman1997stabilizer}
Daniel Gottesman.
\newblock {\em Stabilizer codes and quantum error correction}.
\newblock California Institute of Technology, 1997.

\bibitem{guth2014quantum}
Larry Guth and Alexander Lubotzky.
\newblock Quantum error correcting codes and 4-dimensional arithmetic hyperbolic manifolds.
\newblock {\em Journal of Mathematical Physics}, 55(8):082202, 2014.

\bibitem{hastings2013decoding}
Matthew~B. Hastings.
\newblock Decoding in hyperbolic spaces: Quantum {LDPC} codes with linear rate and efficient error correction.
\newblock {\em Quantum Info. Comput.}, 14(13–14):1187–1202, October 2014.

\bibitem{hastings2020fiber}
Matthew~B. Hastings, Jeongwan Haah, and Ryan O'Donnell.
\newblock Fiber bundle codes: Breaking the $n^{1/2}$poly$\log(n)$ barrier for quantum {LDPC} codes.
\newblock page 1276–1288, 2021.

\bibitem{hong2023long}
Yifan Hong, Matteo Marinelli, Adam~M Kaufman, and Andrew Lucas.
\newblock Long-range-enhanced surface codes.
\newblock {\em Physical Review A}, 110(2):022607, 2024.

\bibitem{horsman2012surface}
Clare Horsman, Austin~G Fowler, Simon Devitt, and Rodney Van~Meter.
\newblock Surface code quantum computing by lattice surgery.
\newblock {\em New Journal of Physics}, 14(12):123011, 2012.

\bibitem{kalachev2022linear}
Gleb Kalachev and Sergey Sadov.
\newblock A linear-algebraic and lattice-theoretical look at the cleaning lemma of quantum coding theory.
\newblock {\em Linear Algebra and its Applications}, 649:96--121, 2022.

\bibitem{kaufman2021new}
Tali Kaufman and Ran~J Tessler.
\newblock New cosystolic expanders from tensors imply explicit quantum ldpc codes with {$\Omega(\sqrt{n}\log^k n)$} distance.
\newblock In {\em Proceedings of the 53rd Annual ACM SIGACT Symposium on Theory of Computing}, pages 1317--1329, 2021.

\bibitem{kitaev1997quantum}
A~Yu Kitaev.
\newblock Quantum computations: algorithms and error correction.
\newblock {\em Russian Mathematical Surveys}, 52(6):1191--1249, 1997.

\bibitem{kitaev2003fault}
A~Yu Kitaev.
\newblock Fault-tolerant quantum computation by anyons.
\newblock {\em Annals of Physics}, 303(1):2--30, 2003.

\bibitem{knill1998resilient}
Emanuel Knill, Raymond Laflamme, and Wojciech~H Zurek.
\newblock Resilient quantum computation: error models and thresholds.
\newblock In {\em Proceedings of the Royal Society of London A: Mathematical, Physical and Engineering Sciences}, volume 454, pages 365--384. The Royal Society, 1998.

\bibitem{kovalev2013quantum}
Alexey~A Kovalev and Leonid~P Pryadko.
\newblock Quantum kronecker sum-product low-density parity-check codes with finite rate.
\newblock {\em Physical Review A—Atomic, Molecular, and Optical Physics}, 88(1):012311, 2013.

\bibitem{kubica2015universal}
Aleksander Kubica and Michael~E Beverland.
\newblock Universal transversal gates with color codes: A simplified approach.
\newblock {\em Physical Review A}, 91(3):032330, 2015.

\bibitem{kurpiers2018deterministic}
Philipp Kurpiers, Paul Magnard, Theo Walter, Baptiste Royer, Marek Pechal, Johannes Heinsoo, Yves Salath{\'e}, Abdulkadir Akin, Simon Storz, J-C Besse, et~al.
\newblock Deterministic quantum state transfer and remote entanglement using microwave photons.
\newblock {\em Nature}, 558(7709):264--267, 2018.

\bibitem{leung2019deterministic}
N~Leung, Y~Lu, S~Chakram, RK~Naik, N~Earnest, R~Ma, K~Jacobs, AN~Cleland, and DI~Schuster.
\newblock Deterministic bidirectional communication and remote entanglement generation between superconducting qubits.
\newblock {\em npj Quantum Information}, 5(1):18, 2019.

\bibitem{leverrier2022quantum}
Anthony Leverrier and Gilles Z{\'e}mor.
\newblock Quantum tanner codes.
\newblock In {\em 2022 IEEE 63rd Annual Symposium on Foundations of Computer Science (FOCS)}, pages 872--883. IEEE, 2022.

\bibitem{li2024transform}
Xingjian Li, Ting-Chun Lin, and Min-Hsiu Hsieh.
\newblock Transform arbitrary good quantum ldpc codes into good geometrically local codes in any dimension.
\newblock {\em arXiv preprint arXiv:2408.01769}, 2024.

\bibitem{lin2023geometrically}
Ting-Chun Lin, Adam Wills, and Min-Hsiu Hsieh.
\newblock Geometrically local quantum and classical codes from subdivision.
\newblock {\em arXiv preprint arXiv:2309.16104}, 2023.

\bibitem{linke2017experimental}
Norbert~M Linke, Dmitri Maslov, Martin Roetteler, Shantanu Debnath, Caroline Figgatt, Kevin~A Landsman, Kenneth Wright, and Christopher Monroe.
\newblock Experimental comparison of two quantum computing architectures.
\newblock {\em Proceedings of the National Academy of Sciences}, 114(13):3305--3310, 2017.

\bibitem{londe2019golden}
Vivien Londe and Anthony Leverrier.
\newblock Golden codes: quantum ldpc codes built from regular tessellations of hyperbolic 4-manifolds.
\newblock {\em Quantum Information \& Computation}, 19(5\&6), 2019.

\bibitem{mccurley1984explicit}
Kevin~S McCurley.
\newblock Explicit estimates for the error term in the prime number theorem for arithmetic progressions.
\newblock {\em Mathematics of computation}, 42(165):265--285, 1984.

\bibitem{monroe2014large}
C~Monroe, R~Raussendorf, A~Ruthven, KR~Brown, P~Maunz, L-M Duan, and J~Kim.
\newblock Large-scale modular quantum-computer architecture with atomic memory and photonic interconnects.
\newblock {\em Physical Review A}, 89(2):022317, 2014.

\bibitem{murali2020architecting}
Prakash Murali, Dripto~M Debroy, Kenneth~R Brown, and Margaret Martonosi.
\newblock Architecting noisy intermediate-scale trapped ion quantum computers.
\newblock In {\em 2020 ACM/IEEE 47th Annual International Symposium on Computer Architecture (ISCA)}, pages 529--542. IEEE, 2020.

\bibitem{alon2016probabilistic}
Joel H.~Spencer Noga~Alon.
\newblock {\em The Probabilistic Method, 4th Edition}.
\newblock Wiley Series in Discrete Mathematics and Optimization. John Wiley \& Sons, 4ed. edition, 2016.

\bibitem{panteleev2020quantum}
Pavel Panteleev and Gleb Kalachev.
\newblock Quantum ldpc codes with almost linear minimum distance.
\newblock {\em IEEE Transactions on Information Theory}, 68(1):213--229, 2021.

\bibitem{panteleev2021asymptotically}
Pavel Panteleev and Gleb Kalachev.
\newblock Asymptotically good quantum and locally testable classical ldpc codes.
\newblock In {\em Proceedings of the 54th Annual ACM SIGACT Symposium on Theory of Computing}, pages 375--388, 2022.

\bibitem{pattison2023hierarchical}
Christopher~A Pattison, Anirudh Krishna, and John Preskill.
\newblock Hierarchical memories: Simulating quantum ldpc codes with local gates.
\newblock {\em arXiv preprint arXiv:2303.04798}, 2023.

\bibitem{periwal2021programmable}
Avikar Periwal, Eric~S Cooper, Philipp Kunkel, Julian~F Wienand, Emily~J Davis, and Monika Schleier-Smith.
\newblock Programmable interactions and emergent geometry in an array of atom clouds.
\newblock {\em Nature}, 600(7890):630--635, 2021.

\bibitem{portnoy2023local}
Elia Portnoy.
\newblock Local quantum codes from subdivided manifolds.
\newblock {\em arXiv preprint arXiv:2303.06755}, 2023.

\bibitem{shor1996fault}
Peter~W Shor.
\newblock Fault-tolerant quantum computation.
\newblock In {\em Proceedings of 37th Conference on Foundations of Computer Science}, pages 56--65. IEEE, 1996.

\bibitem{tillich2014quantum}
Jean-Pierre Tillich and Gilles Z{\'e}mor.
\newblock Quantum {LDPC} codes with positive rate and minimum distance proportional to the square root of the blocklength.
\newblock {\em IEEE Transactions on Information Theory}, 60(2):1193--1202, 2014.

\bibitem{williamson2023layer}
Dominic~J Williamson and Nou{\'e}dyn Baspin.
\newblock Layer codes.
\newblock {\em arXiv preprint arXiv:2309.16503}, 2023.

\bibitem{zhong2021deterministic}
Youpeng Zhong, Hung-Shen Chang, Audrey Bienfait, {\'E}tienne Dumur, Ming-Han Chou, Christopher~R Conner, Joel Grebel, Rhys~G Povey, Haoxiong Yan, David~I Schuster, et~al.
\newblock Deterministic multi-qubit entanglement in a quantum network.
\newblock {\em Nature}, 590(7847):571--575, 2021.

\end{thebibliography}
\end{document}